\documentclass[12pt]{article}
\usepackage{array}
 \usepackage{amsmath,amssymb}
 \usepackage{amsthm}
 \usepackage{bm}
 \usepackage{latexsym}
 \usepackage{CJK}
 \usepackage{multirow}
 \usepackage{caption}
 \usepackage{graphicx,color}
 \usepackage{amssymb,amsfonts}
 \usepackage{hyperref}
 \usepackage[english]{babel}
 \usepackage{times}
 \usepackage[T1]{fontenc}
 \usepackage{enumerate}
\usepackage{geometry}
\usepackage{float}
\usepackage{appendix}
\usepackage{booktabs}
\usepackage{endnotes}
\usepackage{mathtools}
\usepackage{amscd}
\usepackage{bm}
\usepackage{multirow}
\usepackage{booktabs}
\usepackage{graphicx}
\usepackage{amsmath}
\DeclareMathOperator{\sign}{sign}
\usepackage[scaled]{beramono}
\usepackage[T1]{fontenc}
\usepackage[numbers]{natbib}
\usepackage{url}
\usepackage{hyperref}
\usepackage{setspace}
\usepackage{pinlabel}
\usepackage{longtable}
\newtheorem{thm}{Theorem}

\newtheorem{prop}{Proposition}

\newtheorem{condition}{Condition}

\makeatletter
\newcommand*\bigcdot{\mathpalette\bigcdot@{.5}}
\newcommand*\bigcdot@[2]{\mathbin{\vcenter{\hbox{\scalebox{#2}{$\m@th#1\bullet$}}}}}
\makeatother

\begin{document}
\newgeometry{top=1in,bottom=1in, right=1in, left=1in}
	\thispagestyle{empty}
		\title{Integrating multi-source block-wise missing data in model selection}

		\author{Fei Xue and Annie Qu\footnote{Fei Xue is Postdoc Researcher, Department of Biostatistics, Epidemiology and Informatics, University of Pennsylvania, Philadelphia, PA 19104 (E-mail: Fei.Xue@pennmedicine.upenn.edu). Annie Qu is Professor, Department of Statistics, University of California Irvine, Irvine, CA 92697 (E-mail: aqu2@uci.edu).}}
		\date{ }
		\maketitle

		\vspace{-4mm}
	\begin{abstract}  
		For multi-source data, blocks of variable information from certain sources are likely missing. 
		Existing methods for handling missing data do not take structures of block-wise missing data into consideration. In this paper, we propose a Multiple Block-wise Imputation (MBI) approach, which incorporates imputations based on both complete and incomplete observations. Specifically, for a given missing pattern group, the imputations in MBI 
		incorporate more samples from groups with fewer observed variables in addition to the group with complete observations. We propose to construct estimating equations based on all available information, and integrate informative estimating functions to achieve efficient estimators. We show that the proposed method has estimation and model selection consistency under both fixed-dimensional and high-dimensional settings. Moreover, the proposed estimator is asymptotically more efficient than the estimator based on a single imputation from complete observations only. In addition, the proposed method is not restricted to missing completely at random.
		Numerical studies and ADNI data application confirm that the proposed method outperforms existing variable selection methods under various missing mechanisms. 
%
%

				\noindent\textbf{Key words:} ADNI, data integration, dimension reduction, generalized method of moments, informative missing, missing at random
		\end{abstract}
		
\restoregeometry


		\newpage

\section{Introduction}\label{intro}
	
We encounter multi-source or multi-modality data frequently in many real data applications. For example, the Alzheimer's Disease Neuroimaging Initiative (ADNI) data involve multi-site longitudinal observational data from elderly individuals with normal cognition (NC), mild cognitive impairment (MCI), or Alzheimer's Disease (AD) [\citealp{mueller2005ways}, \citealp{mueller2005alzheimer}]. The ADNI data contain multi-source measurements: magnetic resonance imaging (MRI), florbetapir-fluorine-$18$ (AV-$45$) positron emission tomography (PET) imaging, fludeoxyglucose F $18$ (FDG) PET imaging, biosamples, gene expression, and demographic information. Such multi-source data are also common for electronic medical record (EMR) systems adopted by most health care and medical facilities  nowadays, which contain diverse-source patient information, e.g., demographics, medication status,  laboratory tests,  medical imaging and text notes.


However,  blocks of variable information could be completely missing  as   there might be  no need or it might be infeasible to collect certain sources of information given other known  variables. E.g., patients might be either too healthy or too ill.  For EMR systems, it could be due to lack of information exchange or common practice  between different  medical facilities \citep{madden2016missing}. 
Block missing variables cause a large fraction of subjects with certain sources missing, which could lead to biased parameter estimation and inconsistent feature selection. 
Therefore, it is important to fully integrate data from all complementary sources to improve model  prediction and variable selection.

The most common approach for handling missing data is to perform complete-case analysis which removes observations with missing values and only utilizes the complete cases. However, the complete-case method produces biased estimates when the missing is not completely at random. The inverse probability weighting method [\citealp{horvitz1952generalization}] is able to reduce this bias under missing at random mechanism via re-weighting the complete observations [\citealp{seaman2013review}, \citealp{sun2018inverse}]; nevertheless, incomplete observations are still not fully utilized. In real applications, such as the ADNI data, removing incomplete cases could incur a great loss of information since complete cases only account for a small fraction of the data. Alternatively, likelihood-based methods [\citealp{garcia2010variable}, \citealp{ibrahim1999missing}, \citealp{chen2014penalized}] can incorporate all observations. However, this relies on specifying a known distribution which might not be available, and could be computationally intractable if the number of missing variables is large.

Imputation [\citealp{wan2015variable}, \citealp{liu2016variable}] is another widely-used approach to handle missing data. For example, [\citealp{cai2016structured}] propose a structured matrix completion (SMC) method through singular value decomposition to recover a missing block under a low rank approximation assumption. However, the SMC imputes only one missing block at a time. [\citealp{gao2017high}] is capable of imputing all missing values through matrix completion and then apply the adaptive Lasso [\citealp{huang2008adaptive}, \citealp{zou2006adaptive}] to select variables. However, this approach does not guarantee estimation consistency. Alternatively, multiple imputation [\citealp{rubin2004multiple}] (MI) is applicable for conducting variable selection, e.g., [\citealp{chen2013variable}] propose a multiple imputation-least absolute shrinkage and selection operator (MI-LASSO), and adopt the group Lasso [\citealp{yuan2006model}] to detect nonzero covariates. Furthermore, [\citealp{wood2008should}] and [\citealp{wan2015variable}] select variables on combined multiple imputed data. In addition, MI can be combined with bootstrapping techniques [\citealp{heymans2007variable}, \citealp{liu2016variable}, \citealp{long2015variable}]. However, these imputation methods are not effective for block-wise missing data.

Recently, several methods have been developed to target block-wise missing data. E.g., [\citealp{yuan2012multi}] propose an incomplete multi-source feature learning (iMSF) method, which models different missing patterns separately and minimizes a combined loss function. In addition, [\citealp{xiang2014bi}] introduce an incomplete source-feature selection (iSFS) model, utilizing shared parameters across all missing patterns and imposing different weights on different data sources. However, the iSFS is unable to provide coefficient estimation for all samples due to the different weighting strategy. Alternatively, the direct sparse regression procedure using covariance from multi-modality data (DISCOM) [\citealp{yu2019optimal}] estimates the covariance matrices among predictors and between the response and predictors. 
However, the DISCOM only considers missing completely at random, which could be restrictive for missing not completely at random data.




The single regression imputation (SI) method [\citealp{baraldi2010introduction}, \citealp{zhang2016missing}, \citealp{10.1007/978-3-319-25751-8_1}, \citealp{fleet2016initial}, \citealp{saunders2006imputing}] is another popular approach which predicts missing values through regression using observed variables as predictors.
 Suppose that the subjects from multi-source data are divided into groups according to their missing patterns. 
For a group with a given missing block, the SI estimates association between missing variables and observed variables within the group based on complete observations. However, in practice, the complete observations might  only account for a small fraction of the entire data.

To integrate information from the multi-source observed data we propose a Multiple Block-wise Imputation (MBI) approach, incorporating not only the SI based on complete observations but also imputations from incomplete observations. The additional imputations in MBI involve fewer observed variables within a given missing group, but are able to integrate more observations from multiple groups than the SI. Thus, the MBI can improve estimation and model selection especially when the missing rate is high. In addition,  the proposed method aggregates more groups with different missing patterns to impute  missing variables,  which  does not rely on the missing completely at random assumption, and is capable of  handling  missing at random data.


Furthermore, we propose a new multiple block-wise imputation model selection method. Specifically, we propose to construct estimating equations based on all possible missing patterns and imputations, and integrate them through the generalized methods of moments (GMM) [\citealp{hansen1982large}]. 
In theory, we show that the proposed method has estimation and model selection consistency under both fixed-dimensional and high-dimensional settings. Moreover, our estimator is asymptotically more efficient than the SI estimator. Numerical studies and the ADNI data application also confirm that the proposed method outperforms existing variable selection methods for block-wise missing data in missing completely at random, missing at random, and informative missing scenarios.




In general, our work has the following major advantages. First, we are able to integrate the multiple block-wise imputations of all missing pattern groups to improve estimation efficiency and model selection consistency. Second, the proposed method is capable of handling block-wise missing data which might not contain any complete observations, while most traditional methods, including the matrix completion [\citealp{cai2016structured}], require partial subjects to have fully completed observations.

The remainder of the paper is organized as follows. Section \ref{Section2} introduces the background and framework for the block-wise missing problem. In Section \ref{Section3}, we propose the MBI approach incorporating all missing patterns. In Section \ref{Algorithm}, the implementation and algorithm are illustrated.
In Section \ref{SectionTheory}, we establish the theoretical properties of the proposed method.  Sections \ref{SectionSim} and \ref{Real} provide numerical studies through simulations and the ADNI data application.

		\section{Background and Motivation}\label{Section2}

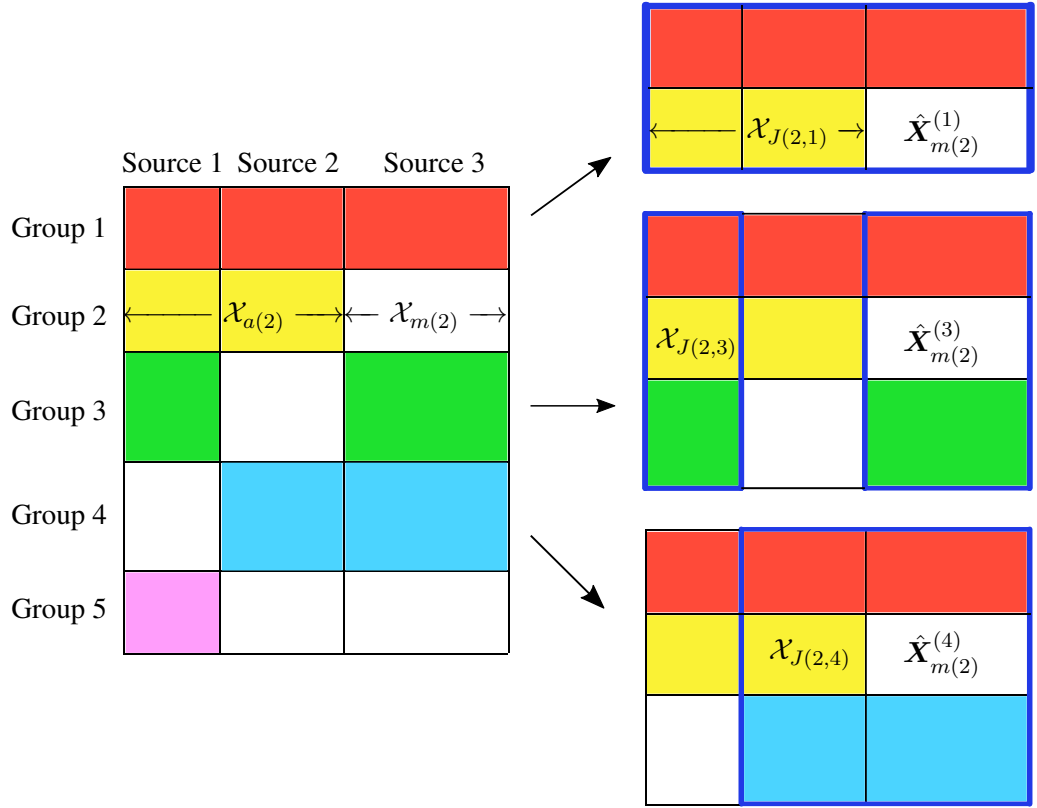
\begin{figure}
		\labellist
		\hair 2pt
		\small
		\pinlabel {Source 1} at 27 352
		\pinlabel {Source 2} at 90 352
		\pinlabel {Source 3} at 170 352
		
		\pinlabel {Group 1} at -35 315
		\pinlabel {Group 2} at -35 267
		\pinlabel {Group 3} at -35 215
		\pinlabel {Group 4} at -35 158
		\pinlabel {Group 5} at -35 107
		
		\pinlabel
		$\xleftarrow{\makebox[1.0cm]{}}\mathcal{X}_{a(2)}\xrightarrow{\makebox[0.45cm]{}}$ at 60 267
		\pinlabel $\xleftarrow{\makebox[0.27cm]{}}\mathcal{X}_{m(2)}\xrightarrow{\makebox[0.3cm]{}}$ at 165 267
		
		\pinlabel$\xleftarrow{\makebox[0.97cm]{}}\mathcal{X}_{J(2,1)}\xrightarrow{\makebox[0.05cm]{}}$ at 346 368
		\pinlabel$\hat{\bm{X}}_{m(2)}^{(1)}$ at 447 368
		\pinlabel $\mathcal{X}_{J(2,3)}$ at 313 253
		\pinlabel$\hat{\bm{X}}_{m(2)}^{(3)}$ at 447 253
		\pinlabel $\mathcal{X}_{J(2,4)}$ at 375 83
		\pinlabel$\hat{\bm{X}}_{m(2)}^{(4)}$ at 447 83
		\endlabellist
		\centering
		\includegraphics[scale=0.7]{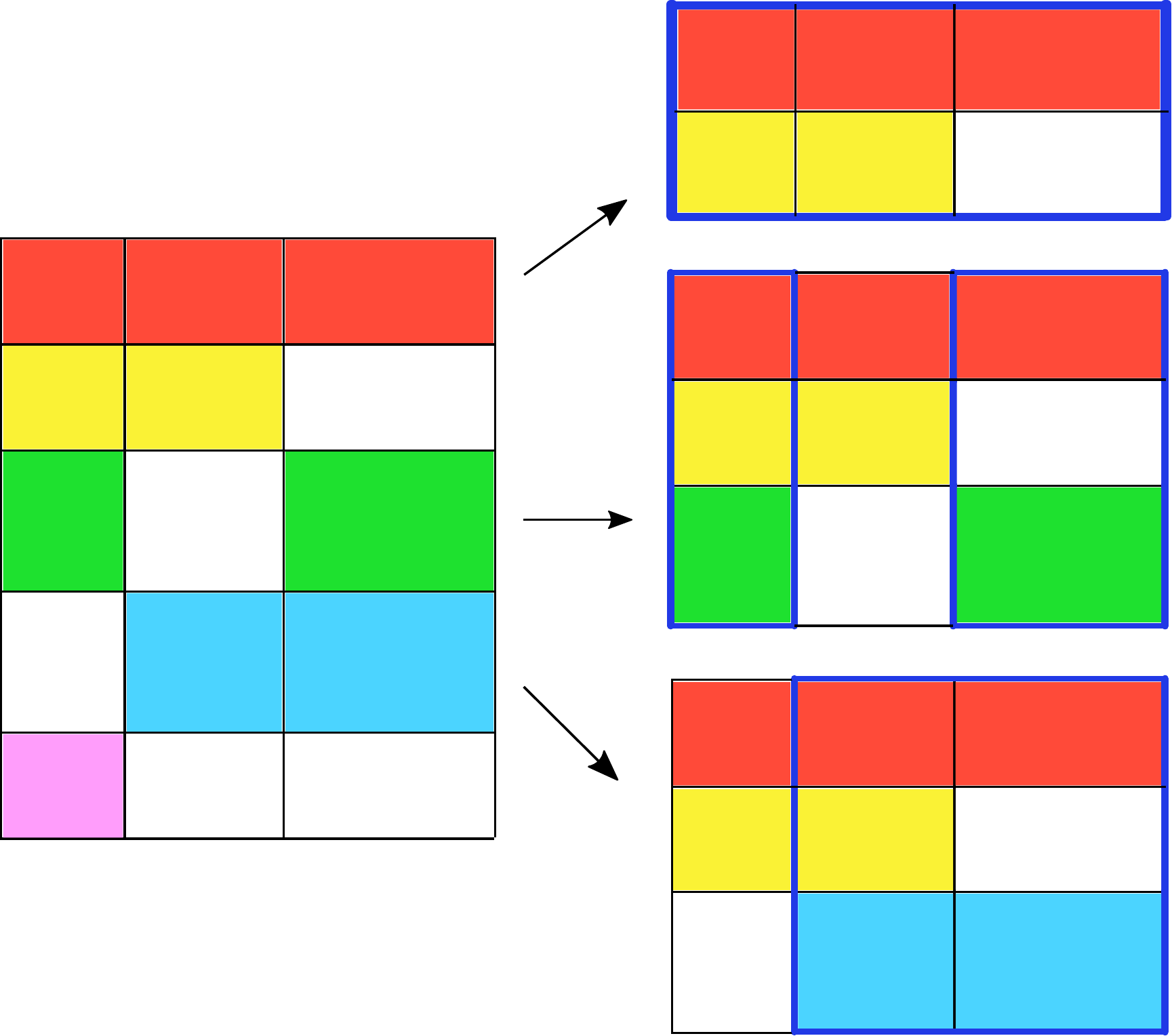}
		\caption{Left: Missing patterns for block-wise missing data. Each white area represents a missing block, while the colored ones represent observed blocks from different missing patterns. Right: Multiple block-wise  imputations for the missing block in Group $2$.}\label{missingpattern}
	\end{figure}

	In this section, we introduce the framework for the block-wise missing problem. 
	Let $\bm{y} = (y_1, \dots, y_n)^T$ be the response variable, and $\bm{X}=(X_{ij})$ be the $N\times p$ design matrix. Suppose that all the samples are drawn independently from a random vector $\bm{\mathcal{X}}=(X_1, X_2, \dots, X_p)$, whose covariance matrix $\bm{C}=(c_{ij})$ is positive definite. Then, for any $1\le i \le N$ and $1\le j \le p$, $X_{ij}$ represents the $i$-the sample of the $j$-th covariate. Suppose that all the covariates in $\bm{X}$ are from $S$ sources. 
	Figure \ref{missingpattern} illustrates a setting with three sources.


	We divide samples $\bm{X}$ into $R$ disjoint groups based on the missing patterns across all sources, where $\bm{x}_i$, the $i$-th row of $\bm{X}$, is in the $r$-th group if $ i \in \mathcal{H}(r)$, and $\mathcal{H}(r)$ is an index set of samples. For any $1\le r\le R$, let $a(r)$ and $m(r)$ be the index sets of the observed covariates and missing covariates corresponding to the $r$-th group, respectively, and obviously, $\bigcup\limits_{r=1}^{R}a(r)=\{1,\dots,p\}$. 
	Then, $\bm{\mathcal{X}}_{a(r)}$ and $\bm{\mathcal{X}}_{m(r)}$ represent observed variables and missing variables in the $r$-th group, respectively. In addition, let $\mathcal{G}(r)$ be the index set of the groups where missing variables $\bm{\mathcal{X}}_{m(r)}$ and variables in at least one of the other sources are observed. If there are no missing values in the $r$-th group, let $\mathcal{G}(r)=\{r\}$, a completely observed dataset. We assume that $\mathcal{G}(r)$ is nonempty containing $M_r=|\mathcal{G}(r)|$ elements for $1\le r\le R$. Note that this assumption does not imply that the data must contain complete observations, since $\mathcal{G}(r)$ could contain a group which is not a complete case group but contains observed values of variables $\bm{\mathcal{X}}_{m(r)}$ and of partial variables in $\bm{\mathcal{X}}_{a(r)}$.

	For illustration, the design matrix on the left of Figure \ref{missingpattern} consists of $3$ sources which are partitioned into $5$ missing pattern groups, where each white area represents a missing block and the colored ones represent observed blocks in different groups. For example, $\mathcal{H}(2)$ refers to samples in Group $2$ and $\bm{\mathcal{X}}_{m(2)}$ refers to missing covariates in Group $2$. Since Groups $1$, $3$ and $4$ contain observed values of $\bm{\mathcal{X}}_{m(2)}$ and covariates in Source $2$ or $3$, $\mathcal{G}(2)=\{1,3,4\}$ and $M_2=3$. If we remove Group $1$ (the complete case group) and Group $5$, 
		 the $\mathcal{G}(2)$ is still non-empty, which is also true for $\mathcal{G}(3)$ and $\mathcal{G}(4)$. 

	
	
	We consider the following linear model
	\begin{equation}\label{Lmodel}
	\bm{y}=\bm{X}\bm{\beta}^0 + \bm{\varepsilon},
	\end{equation}
	where $\bm{\beta}^0=(\beta_1^0, \dots, \beta_p^0)^T$ is the true coefficient vector corresponding to all covariates and $\bm{\varepsilon}\sim N(\bm{0}, \sigma_{\varepsilon}^2\bm{I}_N)$ represents an error term independent of $\bm{X}$. 
	We assume that the model is sparse; that is, most of the true coefficients are zero. 
	Let $A_1=\{j : \beta_j^0\ne0\}$ and $A_2=\{j : \beta_j^0=0\}$ be index sets corresponding to relevant and irrelevant covariates, respectively. We also let $q=|A_1|$ be the total number of relevant covariates. In the supplementary material, we provide a table of all notations for convenience.

	The likelihood-based approaches [\citealp{garcia2010variable}] typically formulate likelihood based on completely observed variables. However, it is likely that no covariate is completely observed under the block-wise missing structure. Alternatively,  [\citealp{yuan2012multi}] construct a model for each missing pattern separately and use observed variables within each missing pattern as predictors. For instance, for Group $2$ in Figure \ref{missingpattern}, the above method treats the covariates in Sources $1$ and $2$ as predictors and ignores information from Source $3$. However, Source $3$ covariates could be incorporated as well, since they are relevant to the response variable. 
	

Traditional imputation methods [\citealp{zhang2016missing}, \citealp{10.1007/978-3-319-25751-8_1}, \citealp{fleet2016initial}] impute missing values in Group $2$ based on the associations between missing and observed variables obtained from complete observations in Group $1$, while
	samples in Groups $3$ and $4$ are not utilized. However, Groups $3$ and $4$, also containing values from Source $3$, can provide additional information in imputing missing variables $\bm{\mathcal{X}}_{m(r)}$ through correlations with other covariates.
	This is especially useful when completely observed subjects are scarce. In the following section, we propose a new imputation approach to fully utilize information not only from the group with complete cases but also from other groups. 
	
	
		

%
			
			\section{Method} \label{Section3}

\subsection{ Multiple Block-wise Imputation}\label{MBI}

In this subsection, we propose a multiple block-wise imputation approach which can utilize more observed information from incomplete case groups than traditional imputation methods.
Specifically, for a given Group $r$ with missing values of $\bm{\mathcal{X}}_{m(r)}$, 
each of the $\mathcal{G}(r)$ groups contains observed values corresponding to missing $\bm{\mathcal{X}}_{m(r)}$, and also observed values corresponding to a subset of observed $\bm{\mathcal{X}}_{a(r)}$. Therefore, we can predict missing values in the $r$-th group with $M_r=|\mathcal{G}(r)|$ ways to borrow information from all the groups in $\mathcal{G}(r)$, instead of using a complete case group only.

More specifically, for each $k \in \mathcal{G}(r)$, let $J(r,k)=a(r)\cap a(k)$ be an index set of covariates which are observed in Groups $r$ and $k$. For each $j \in m(r)$, we estimate $E(X_j|\bm{\mathcal{X}}_{J(r,k)})$ utilizing all the groups containing observed values of both $X_j$ and $\bm{\mathcal{X}}_{J(r,k)}$, and then impute missing values for $X_j$ in the $r$-th group using association information in the conditional expectation. Let $\hat{\bm{X}}_{m(r)}^{(k)}$ represent the imputation for all missing values in Group $r$. The proposed multiple imputations approach is referred to as Multiple Block-wise Imputation (MBI). We illustrate the MBI with an example in Figure \ref{missingpattern}. For Group $2$, 
covariates observed in both Group $2$ and a group in $\mathcal{G}(2)=\{1,3,4\}$ are indexed by $J(2,1)=B_1\cup B_2$, $J(2,3)=B_1$, and $J(2,4)=B_2$, respectively, where $B_k$ is an index set of covariates from Source $k$ for $k=1,2,3$.

The traditional imputation methods, such as the SI, only utilize observed values in Group $2$ and Group $1$ to impute the missing values in Group $2$, namely $\hat{\bm{X}}_{m(2)}^{(1)}$, as shown on the top right of Figure \ref{missingpattern}. In contrast, the proposed method can incorporate more information from Groups $3$ and $4$ in addition to Groups $1$ and $2$, and impute the missing values in Group $2$ using three different blocks of observed variables. Namely, we estimate $E(\bm{\mathcal{X}}_{B_3} | \bm{\mathcal{X}}_{B_1\cup B_2})$, $E(\bm{\mathcal{X}}_{B_3} | \bm{\mathcal{X}}_{B_1})$ and $E(\bm{\mathcal{X}}_{B_3} | \bm{\mathcal{X}}_{B_2})$ based on Group $1$, Groups $1$ and $3$, and Groups $1$ and $4$, respectively. We then impute the missing values via the above three estimated conditional expectations and the observed information in Group $2$. 
	 Compared with the SI, the proposed MBI incorporates additional imputed values $\hat{\bm{X}}_{m(2)}^{(3)}$ and $\hat{\bm{X}}_{m(2)}^{(4)}$ via $E(\bm{\mathcal{X}}_{B_3} | \bm{\mathcal{X}}_{B_1})$ and $E(\bm{\mathcal{X}}_{B_3} | \bm{\mathcal{X}}_{B_2})$, where the estimation involves more observed samples than the SI approach.
In particular, when estimating conditional expectation for the imputations, we aggregate subjects from different missing pattern groups, 
which can diminish the influence of specific missing patterns of covariates.

	\vspace{-3mm}
\subsection{Integration of MBI}


	In this subsection, we propose to integrate information from all available sources and multiple block-wise imputations.
	Specifically, we construct estimating functions for each group according to its missing pattern. For a given Group $r$ containing missing values and $k\in\mathcal{G}(r)$, since missing $\bm{\mathcal{X}}_{m(r)}$ are estimated through $E(\bm{\mathcal{X}}_{m(r)} | \bm{\mathcal{X}}_{J(r,k)})$  which is a projection onto $\bm{\mathcal{X}}_{J(r,k)}$,
	the covariates $\bm{\mathcal{X}}_{J(r,k)}$ are uncorrelated with residuals of the projection $\bm{\mathcal{X}}_{m(r)} - E(\bm{\mathcal{X}}_{m(r)} | \bm{\mathcal{X}}_{J(r,k)})$.
	Therefore, for each $j \in J(r,k)$,
	\begin{eqnarray*}\label{CEequality1}
	&&E\left[ X_{j} \left\{y-\bm{\mathcal{X}}_{a(r)}\bm{\beta}^0_{a(r)} - E(\bm{\mathcal{X}}_{m(r)} | \bm{\mathcal{X}}_{J(r,k)})\bm{\beta}^0_{m(r)}\right\}\right] \notag\\
	&=&E\left(X_j \ \varepsilon\right)+E\left[ X_{j} \left\{\bm{\mathcal{X}}_{m(r)} - E(\bm{\mathcal{X}}_{m(r)} | \bm{\mathcal{X}}_{J(r,k)})\right\}\right]\bm{\beta}^0_{m(r)}=0,
	\end{eqnarray*}
 where 
	$\bm{\beta}^0_{a(r)}$ and $\bm{\beta}^0_{m(r)}$ denote the true coefficients of $\bm{\mathcal{X}}_{a(r)}$ and $\bm{\mathcal{X}}_{m(r)}$, respectively.
	In addition, for any $j \in m(r)$, since $E(X_{j}|\bm{\mathcal{X}}_{J(r,k)})$ is a function of $\bm{\mathcal{X}}_{J(r,k)}$,
	\begin{eqnarray*}\label{CEequality2}
	&&E\left[ E(X_{j}|\bm{\mathcal{X}}_{J(r,k)}) \left\{y-\bm{\mathcal{X}}_{a(r)}\bm{\beta}^0_{a(r)} - E(\bm{\mathcal{X}}_{m(r)} | \bm{\mathcal{X}}_{J(r,k)})\bm{\beta}^0_{m(r)}\right\}\right] \notag\\
	&=&E\left\{ E(X_{j}|\bm{\mathcal{X}}_{J(r,k)})\ \varepsilon\right\}+E\left[ E(X_{j}|\bm{\mathcal{X}}_{J(r,k)}) \left\{\bm{\mathcal{X}}_{m(r)} - E(\bm{\mathcal{X}}_{m(r)} | \bm{\mathcal{X}}_{J(r,k)})\right\}\right]\bm{\beta}^0_{m(r)}=0. 
	\end{eqnarray*}
	Note that $J(r,k) \cup m(r)=\left\{a(r)\cap a(k)\right\} \cup m(r)=a(k)$ since $m(r)\subset a(k)$ for $k \in \mathcal{G}(r)$.

	Thus, we construct estimating functions corresponding to observed covariates in the $k$-th group using imputed values $\hat{\bm{X}}_{m(r)}^{(k)}$. 
	In general, for each $ i \in \mathcal{H}(r)$, let $\bm{x}_i^{(k)}=(X_{i1}^{(k)}, \dots, X_{ip}^{(k)})$ be the $i$-th imputed sample based on Group $k$, 
	where $X_{ij}^{(k)} = X_{ij}$ if the $j$-th covariate is observed in the sample $\bm{x}_i$, otherwise $X_{ij}^{(k)}$ is an imputed value of $X_{ij}$ in $\hat{\bm{X}}_{m(r)}^{(k)}$.
	The estimating functions for the imputed samples $\bm{x}_i^{(k)}$ in Group $r$ are	
	\begin{eqnarray*}
		\bm{g}_i^{(r , k)} (\bm{\beta})=
		\frac{\partial \mu_i^{(k)} (\bm{\beta})}{\partial \bm{\beta}_{a(k)}} 
	\left\{y_i-\mu_i^{(k)} (\bm{\beta})\right\} = \left\{\bm{z}_i^{(k)}\right\}^T \left\{y_i-\mu_i^{(k)} (\bm{\beta})\right\} \ \ \ \ \text{ for } i\in \mathcal{H}(r),
	\end{eqnarray*}
	where 
	$\bm{z}_i^{(k)}$ is a sub-vector of $\bm{x}_i^{(k)}$ consisting of $X_{ij}^{(k)}$ for $j\in a(k)$, 
	$\partial \mu_i^{(k)} /\partial \bm{\beta}_{a(k)}$ is the derivative of $\mu_i^{(k)} (\bm{\beta}) = \bm{x}_i^{(k)} \bm{\beta}$ with respect to $\bm{\beta}_{a(k)}$, and $\bm{\beta}_{a(k)}$ is the coefficient vector corresponding to $\bm{\mathcal{X}}_{a(k)}$. 

To integrate information from all available missing patterns and imputations, we propose an aggregated vector of estimating functions:
	\vspace{-3mm}
\begin{equation}\label{AllEF}
	\vspace{-3mm}
\bm{g}(\bm{\beta})=(\{\bm{g}^{(1)} (\bm{\beta})\}^T,\dots, \{\bm{g}^{(R)} (\bm{\beta})\}^T)^T,
\end{equation}
where
	\vspace{-3mm}
\begin{equation*}
	\vspace{-3mm}
\bm{g}^{(r)}=\frac{1}{n_r}\sum_{i\in\mathcal{H}(r)}\bm{g}_i^{(r)} (\bm{\beta}),
\end{equation*}
$n_r$ is the number of samples from the $r$-th group, and $\bm{g}_i^{(r)} (\bm{\beta})$ is a vector consisting of $\bm{g}_i^{(r, k)} (\bm{\beta})$ for $k\in\mathcal{G}(r)$.
	If the $r$-th group only has complete observations, then $\mathcal{G}(r)=\{r\}$, $M_r=1$ and 
		\vspace{-3mm}
\begin{equation*}
	\vspace{-3mm}
\bm{g}_i^{(r)}(\bm{\beta})=\bm{g}_i^{(r, r)}(\bm{\beta})= \bm{x}_{i}^T \left\{y_i-\bm{x}_i\bm{\beta}
\right\} \ \ \ \ \ \  \ \ \  \text{ for } i\in \mathcal{H}(r).
\end{equation*} 
	
	Note that the total number of equations exceeds the number of coefficient parameters, and estimating functions from groups with fewer missing variables or more accurate imputations tend to have smaller variance. 
	To combine all the estimating functions in $\bm{g}(\bm{\beta})$, 
	we estimate coefficients $\bm{\beta}$ through the penalized generalized method of moments \citep{caner2009lasso} which minimizes 
	\vspace{-3mm}
	\begin{equation}\label{nonsingular}
	\vspace{-3mm}
	f(\bm{\beta})=\{\bm{g}(\bm{\beta})\}^T\bm{W}(\bm{\beta})^{-1}\bm{g}(\bm{\beta})+\sum\limits_{j=1}^{p} p_{\lambda}(|\beta_j|),
	\end{equation}
	where 
		\vspace{-3mm}
	 \begin{equation*}
	 	\vspace{-3mm}
	 \bm{W}(\bm{\beta})=diag\left\{\frac{1}{n_1}\sum_{i\in\mathcal{H}(1)}\bm{g}_i^{(1)}(\bm{\beta})\{\bm{g}_i^{(1)}(\bm{\beta})\}^T, \dots, \frac{1}{n_R}\sum_{i\in\mathcal{H}(R)}\bm{g}_i^{(R)}(\bm{\beta})\{\bm{g}_i^{(R)}(\bm{\beta})\}^T \right\}
	 \end{equation*}
	is the sample covariance matrix of $\bm{g}(\bm{\beta})$, and $p_{\lambda}(\cdot)$
	is a penalty function with tuning parameter $\lambda$. 
	In this paper, we choose the SCAD penalty due to its oracle properties \citep{fan2001variable}. The sample covariance matrix $\bm{W}(\bm{\beta})$ is a block diagonal matrix since estimating functions are formulated based on different missing patterns. 
	However, $\bm{W}(\bm{\beta})$ could be singular or close to singular due to overlapping information in imputations, or due to a large number of estimating functions compared to a relatively small sample size. For example, as illustrated in Figure \ref{missingpattern}, the observed values of Source $1$ covariates in Group $2$ are utilized in the estimation of both $\hat{\bm{X}}_{m(2)}^{(1)}$ and $\hat{\bm{X}}_{m(2)}^{(3)}$.

	\vspace{-3mm}
	\subsection{Solving the singularity issue of estimating equations}

	To solve the singularity issue of estimating equations, we reduce the dimension of $\bm{g}^{(r)}$ for $r=1,\dots,R$,
	through combining informative estimating equations, e.g., utilizing the first several largest principle components (PCs) [\citealp{wold1987principal}, \citealp{cho2015efficient}]. Specifically,  we divide the estimating functions in $\bm{g}^{(r)}$ into two parts $\bm{g}_{(1)}^{(r)}$ and $\bm{g}_{(2)}^{(r)}$, where $\bm{g}_{(1)}^{(r)}$ consists of the functions with the imputation based on complete observations, and $\bm{g}_{(2)}^{(r)}$ contains the remaining estimating functions in $\bm{g}^{(r)}$. We proceed to extract informative principle components from $\bm{g}_{(1)}^{(r)}$ and $\bm{g}_{(2)}^{(r)}$ separately.
	 Let Group $1$ be the complete case group, and $\bm{W}_{11}^{(r)}$ and $\bm{W}_{22}^{(r)}$ be the sample covariance matrices of $\bm{g}_{(1)}^{(r)}$ and $\bm{g}_{(2)}^{(r)}$, respectively. If the dimension of $\bm{g}_{(1)}^{(r)}$ is too large such that $\bm{W}_{11}^{(r)}$ is singular or close to singular, we extract the first $t_1$ principle components $\bm{h}^{(r)}=\bm{U}_1^{(r)}\bm{g}_{(1)}^{(r)}$ from $\bm{g}_{(1)}^{(r)}$, where $\bm{U}_1^{(r)}$ contains $t_1$ eigenvectors of $\bm{W}_{11}^{(r)}$ corresponding to the largest $t_1$ nonzero eigenvalues, and $t_1$ can be selected to retain sufficient information. If $\bm{W}_{11}^{(r)}$ is neither singular nor close to singular, we retain all the estimating functions in $\bm{g}_{(1)}^{(r)}$, and let $\bm{U}_1^{(r)}$ be an identity matrix, that is, $\bm{h}^{(r)}=\bm{g}_{(1)}^{(r)}$.
	
	 We orthogonalize $\bm{g}_{(2)}^{(r)}$ against the $\bm{h}^{(r)}$ to store additional information beyond $\bm{h}^{(r)}$, where $\bar{\bm{g}}_{(2)}^{(r)}=\bm{g}_{(2)}^{(r)}-\bm{V}_{21}^{(r)}\{\bm{V}_{11}^{(r)}\}^{-1}\bm{h}^{(r)}$ consists of orthogonalized estimating functions, $\bm{V}_{11}^{(r)}=\bm{U}_1^{(r)} \bm{W}_{11}^{(r)} \{\bm{U}_1^{(r)}\}^T$,
	and $\bm{V}_{21}^{(r)}$ is the sample covariance matrix between $\bm{g}_{(2)}^{(r)}$ and $\bm{h}^{(r)}$.  Similarly, if the sample covariance of $\bar{\bm{g}}_{(2)}^{(r)}$ is singular or close to singular, we select the first $t_2$ principle components $\bm{U}_2^{(r)}\bar{\bm{g}}_{(2)}^{(r)}$ from the orthogonalized $\bar{\bm{g}}_{(2)}^{(r)}$, where $\bm{U}_2^{(r)}$ contains $t_2$ eigenvectors of the sample covariance matrix of $\bar{\bm{g}}_{(2)}^{(r)}$ corresponding to the largest $t_2$ nonzero eigenvalues.
	Otherwise, we retain all the $\bar{\bm{g}}_{(2)}^{(r)}$, and let $\bm{U}_2^{(r)}$ be an identity matrix. 
	
	Let
	\begin{equation*}
	\bm{U}^{(r)}=	\begin{pmatrix}\bm{U}_1^{(r)}&\bm{0}\\-\bm{U}_2^{(r)}\bm{V}_{21}^{(r)}\{\bm{V}_{11}^{(r)}\}^{-1}\bm{U}_1^{(r)}\ & \bm{U}_2^{(r)}\end{pmatrix}.
	\end{equation*}
	If there is no complete case group or $M_r=1$, then either $\bm{g}_{(1)}^{(r)}$ or $\bm{g}_{(2)}^{(r)}$ is null, and $\bm{U}^{(r)}$ is either $\bm{U}_2^{(r)}$ or $\bm{U}_1^{(r)}$.
	Thus, $\bm{U}^{(r)}\bm{g}^{(r)}$ contains all the essential information from the estimating functions of the $r$-th group,  while solving the singularity issue of the sample covariance matrix. The numbers of principle components $t_1$ and $t_2$ can be tuned through the Bayesian information type of criterion proposed by \citep{cho2015efficient} to capture sufficient information from the estimating functions in (\ref{AllEF}). Consequently, the proposed estimator $\hat{\bm{\beta}}$ is obtained via minimizing 
	\vspace{-3mm}
		\begin{equation}\label{FinalObject}
		\vspace{-3mm}
	f^*(\bm{\beta})=(\bm{Ug})^T(\bm{UW}\bm{U}^T)^{-1}\bm{Ug}+\sum\limits_{j=1}^{p} p_{\lambda}(|\beta_j|),
	\end{equation}
	where $\bm{U}=diag\{\bm{U}^{(1)}, \dots, \bm{U}^{(R)}\}$. In the following section, we also provide an algorithm and implementation strategy of the proposed method.
	


\vspace{-3mm}
\section{Implementation}\label{Algorithm}

In this section, we provide the detailed algorithm for the proposed method. 
The conditional expectations of missing covariates in MBI can be estimated via linear regression models, generalized linear models (GLM) or non-parametric models.
  In this paper, we utilize the GLM \citep{10.2307/2344614} to accommodate not only continuous covariates but also discrete covariates.
  Specifically, for each group $1\le r\le R$, $j\in m(r)$, and $k\in\mathcal{G}(r)$, we adopt the GLM  to predict $E(X_j|\bm{\mathcal{X}}_{J(r,k)})$ if groups containing observed values of both $X_j$ and $\bm{\mathcal{X}}_{J(r,k)}$ have a larger sample size than the number of observed variables $|J(r,k)|$, or adopt the $L_1$-regularized GLM \citep{friedman2010regularization} otherwise.
  To obtain the $L_1$-regularized GLM estimator, we apply the ``glmnet'' package (\resizebox{31ex}{1.4ex}{\href{https://cran.r-project.org/web/packages/glmnet/index.html}{\ttfamily https://cran.r-project.org/web}} \resizebox{28ex}{1.4ex}{\href{https://cran.r-project.org/web/packages/glmnet/index.html}{\ttfamily packages/glmnet/index.html}}) in R. The imputed values in MBI are then computed based on the estimated conditional expectation.
  
  
  To consistently handle singular or non-singular $\bm{W}$\hspace{-0.5mm}, we use $f^*(\bm{\beta})$ in (\ref{FinalObject}) instead of $f(\bm{\beta})$ in (\ref{nonsingular}) as our objective function for all $\bm{W}$\hspace{-0.5mm}. For the sample covariance matrix $\bm{W}^{(r)}$ of $\bm{g}^{(r)}$ from Group $r$, we select the numbers of principle components $t_1$ and $t_2$ corresponding to $\bm{\Omega}$, which is $\bm{W}^{(r)}_{11}$ or $\bm{W}^{(r)}_{22}-\bm{V}^{(r)}_{21}\{\bm{V}^{(r)}_{11}\}^{-1}\{\bm{V}^{(r)}_{21}\}^T$, through minimizing the BIC-type of criterion \citep{cho2015efficient}
  \begin{equation}\label{PC}
  \Psi(t) = \frac{\text{tr}\{\bm{\Omega}-\widetilde{\bm{\Omega}}(t)\}}{\text{tr}\{\bm{\Omega}\}} + t\frac{\log (n_r d)}{n_r d},
  \end{equation}
  where  
$d$ is the dimension of $\bm{\Omega}$. Here, the
  $\widetilde{\bm{\Omega}}(t)=\sum_{j=1}^{t}\lambda_j\bm{v}_j\bm{v}_j^T$ is an approximation of $\bm{\Omega}$ based on the $t$ largest eigenvectors, where $\lambda_j$ is the $j$-th largest eigenvalue of $\bm{\Omega}$, and $\bm{v}_j$ is the eigenvector of $\bm{\Omega}$ corresponding to $\lambda_j$. Since 
   $\text{tr}\{\bm{\Omega}-\widetilde{\bm{\Omega}}(t)\}=\sum_{j=t+1}^{d}\lambda_j$, the minimizer of $\Psi(t)$ is indeed the number of eigenvalues which are larger than $\text{tr}\{\bm{\Omega}\} \log (n_r d)/(n_r d)$.
   
   \begin{figure}
\labellist
\pinlabel
$f^*(\bm{\beta})$ at 5 450
\pinlabel $\beta_{2}$ at 170 85
\pinlabel $\beta_{1}$ at 630 75
\endlabellist
\centering
   	\includegraphics[scale=0.5]{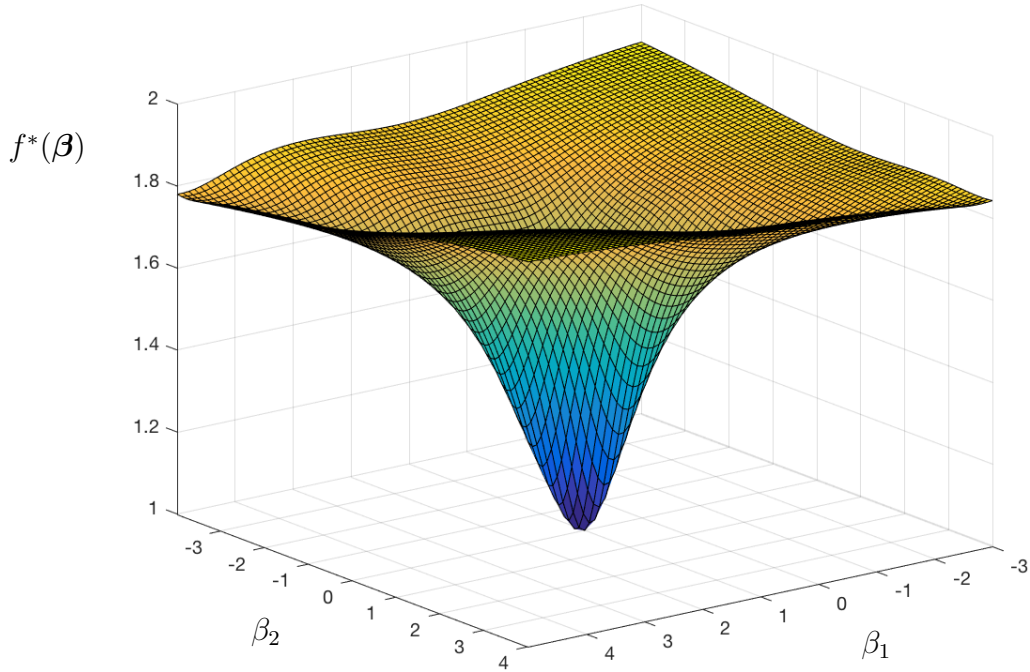}
   	\caption{The objective function $f^*(\bm{\beta})$.}\label{objective}
   \end{figure}

	\begin{table}[htbp]\centering
	\begin{tabular}{c}
		\hline
		\textbf{Algorithm 1} \\
		\hline
		\parbox{14.5cm}{
			\begin{enumerate}
				\item Obtain initial values $\bm{\beta}^{(0)}$ based on complete observations. Set tolerance $\epsilon$, and the tuning parameters $\lambda$ and $a$.
				\item Estimate $\hat{\bm{X}}^{(j)}_{m(r)}$ via the GLM or $L_1$-regularized GLM depending on the sample size for each $r=1, \dots, R$ and $j\in\mathcal{G}(r)$.
				\item At the $k$-th iteration, given $\bm{\beta}^{(k-1)}$ and $\bm{s}_{k-1}$ from the ($k-1$)-th iteration:
				\begin{enumerate} 
					\item Select the number of principle components using (\ref{PC})  if $\bm{W}^{(r)}$ is singular for $r=1,\dots, R$.
					\item Calculate the conjugate direction $\bm{s}_k$ using (\ref{direction}).
					\item  Calculate the step size $\alpha_k$ using (\ref{step}).
					\item Update $\bm{\beta}^{(k)}=\bm{\beta}^{(k-1)}+\alpha_k \bm{s}_k$.
				\end{enumerate}
				
				\item Iterate Step 3 until the convergence criterion is satisfied, e.g., $\underset{1\le j \le p}{\max}\{|\beta_j^{(k)}-\beta_j^{(k-1)}|\}<\epsilon$.
		\end{enumerate}}\\
		\hline
	\end{tabular}
\end{table}

We plot an example of the objective function $f^*(\bm{\beta})$ in Figure \ref{objective} to illustrate the objective function $f^*(\bm{\beta})$ near true coefficients. In this example, there are three sources and four groups with $p_1=p_2=p_3=20$ and $n_1=n_2=n_3=n_4=1000$, where each source contains one relevant predictor with a signal strength of $1$ and the missing patterns are the same as in Groups $1$--$4$ in Figure \ref{missingpattern}. The true coefficients of $\beta_1$ and $\beta_{2}$ are $1$ and $0$, respectively. Figure \ref{objective} shows that $f^*(\bm{\beta})$ has a unique minimizer around the true coefficients.

To obtain the minimizer, we propose to iteratively decrease $f^*(\bm{\beta})$ via the nonlinear conjugate gradient algorithm \citep{dai1999nonlinear} which converges quadratically \citep{cohen1972rate} without requiring the second derivative of the objective function. At the $k$-th iteration, the conjugate direction is 
\vspace{-3mm}
\begin{equation}\label{direction}
\vspace{-3mm}
\bm{s}_k=-\nabla f^*(\bm{\beta}^{(k-1)})+\gamma_{k-1}\bm{s}_{k-1},
\end{equation}
where $\nabla f^*(\bm{\beta}^{(k-1)})$
is the gradient of $f^*(\bm{\beta})$ at $\bm{\beta}=\bm{\beta}^{(k-1)}$, 
\begin{equation*}
\gamma_{k-1}= - \frac{\left\{\nabla f^*(\bm{\beta}^{(k-1)})\right\}^T\nabla f^*(\bm{\beta}^{(k-1)})}{\bm{s}_{k-1}^T \left\{\nabla f^*(\bm{\beta}^{(k-2)})-\nabla f^*(\bm{\beta}^{(k-1)})\right\}},
\end{equation*}
and $\bm{s}_1=-\nabla f^*(\bm{\beta}^{(0)})$. Here, the initial values $\bm{\beta}^{(0)}$ are obtained by performing the Lasso method \citep{tibshirani1996regression} on complete observations, and the gradient is numerically calculated via central differences.
We determine the step size in the conjugate direction $\bm{s}_k$ through a line search:
\vspace{-3mm}
\begin{equation}\label{step}
\vspace{-3mm}
\alpha_k=\underset{\alpha}{\text{argmin}} f^*(\bm{\beta}^{(k-1)}+\alpha \bm{s}_k).
\end{equation} 
We summarize the whole procedure for the implementation of the proposed method in Algorithm 1. Note that estimation of MBI is carried out in Step 2, and the nonlinear conjugate gradient method is performed in Step 3.

To select the tuning parameter $\lambda$ in the penalty function $p_{\lambda} (\cdot)$, we propose a BIC-type criterion (MBI-BIC) as follows:
	\vspace{-3mm}
\begin{equation}\label{MBI-BIC}
	\vspace{-3mm}
\text{MBI-BIC}_{\lambda}=N\cdot\log\left\{RSS(\hat{\bm{\beta}}_{\lambda})/N\right\}+df_{\lambda}\cdot\log (N),
\end{equation}
where $\hat{\bm{\beta}}_{\lambda}$ is the proposed estimator for a given $\lambda$, $df_{\lambda}$ is the number of non-zero estimated coefficients in $\hat{\bm{\beta}}_{\lambda}$, and $RSS(\hat{\bm{\beta}}_{\lambda})=\sum_{r=1}^{R} RSS_r(\hat{\bm{\beta}}_{\lambda})$ is the residual sum of squares from all the missing pattern groups with the $r$-th group
\vspace{-3mm}
\begin{equation*}
\vspace{-3mm}
 RSS_r(\hat{\bm{\beta}}_{\lambda})= \frac{1}{M_r} \sum_{j\in\mathcal{G}(r)} \sum_{i\in\mathcal{H}(r)}\left\{y_i-\mu_i^{(j)}(\hat{\bm{\beta}}_{\lambda})\right\}^2.
\end{equation*}
Here we tune $\lambda$ via the BIC-type criterion instead of $RSS(\hat{\bm{\beta}}_{\lambda})$ to reduce the complexity of model and avoid over-fitting. Compared with the traditional Bayesian information criterion (BIC) \citep{schwarz1978estimating}, the proposed MBI-BIC incorporates additional information from incomplete observations via the MBI. 
We select the optimal tuning parameter $\lambda$ corresponding to the lowest MBI-BIC.

	\vspace{-3mm}
			\section{Theory} \label{SectionTheory}
			
			
			In this section, we provide the theoretical foundation of the proposed method under regularity conditions. In particular, we establish estimation consistency, selection consistency, and asymptotic normality of the proposed estimator. We also show that the proposed MBI leads to more efficient estimation than a single imputation method. Throughout this section, we assume that sources of covariates for each subject are missing completely at random (MCAR) or missing at random (MAR). Let $\xi_i$ be the random group label for the $i$-th subject, that is, $\xi_i=r$ if and only if the subject $i$ is in Group $r$. The MCAR or MAR assumption implies that $\xi_i$ is independent of all covariates or only depends on observed covariates. In addition, we assume that $\xi_i$ for $1\le i \le N$ are independently and identically distributed. 
			
				\vspace{-3mm}
			\subsection{Asymptotic properties for fixed $p$ and $q$}
			
			In this subsection, we assume that $p$ and $q$ are both fixed as $N\to\infty$.
			Let $\bm{G}(\bm{\beta})=(\bm{g}_1(\bm{\beta}), \dots, \linebreak \bm{g}_N(\bm{\beta}))^T$ be the estimating functions from $N$ samples, where $\bm{g}_i(\bm{\beta})=(\bm{g}_i^{(1)}(\bm{\beta})^T, \dots, \bm{g}_i^{(R)}(\bm{\beta})^T)^T$ is a column vector consisting of the $i$-th sample of all available estimating functions with $\bm{g}_i^{(r)}(\bm{\beta})=\bm{0}$ if $i\notin \mathcal{H}(r)$ for $1\le r\le R$. 
			We require the following regularity conditions:
			
			
			
	\begin{condition}\label{C1}
		For any $1\le r \le R$, $k\in \mathcal{G}(r)$, $l  \in m(r)$, and $j \in a(k)$, there exists a sequence $\zeta_N$
			such that $1/\zeta_N=O(1)$, $\zeta_N=o(\sqrt{N})$, 
		\begin{equation}\label{imputeCondition1}
		\frac{1}{N}\sum\limits_{i=1}^N I(\xi_i=r) \hat{Z}_{ij}\left\{E(X_{il}|\bm{X}_{i J(r,k)})-\hat{E}(X_{l}|\bm{X}_{i J(r,k)})\right\}=O_p\left(\frac{\zeta_N}{\sqrt{N}}\right),
		\end{equation}
		\begin{equation}\label{imputeCondition2}
		\frac{1}{\sqrt{N}}\sum\limits_{i =1}^N I(\xi_i=r) \bar{\varepsilon}_i \left\{E(X_{il}|\bm{X}_{iJ(r,k)})-\hat{E}(X_{l}|\bm{X}_{iJ(r,k)})\right\}=O_p\left(\frac{\zeta_N}{\sqrt{N}}\right),
		\end{equation}
		and
		$$E(X_j^4)<\infty, E(X_l^4)<\infty, E(\varepsilon^4)<\infty, E\left\{\hat{E}(X_{l}|\bm{\mathcal{X}}_{J(r,k)})\right\}^4<\infty,$$
		where $I(\cdot)$ is an indicator function, $\bm{X}_{i J(r,k)}$ is a vector consisting of samples $X_{iv}$ for all $v\in J(r,k)$, and $\hat{E}(X_{l}|\bm{X}_{i J(r,k)})$ is an estimator of $E(X_{l}|\bm{\mathcal{X}}_{J(r,k)}=\bm{X}_{i J(r,k)})$. 
			$\hat{Z}_{ij}=\hat{E}(\bm{X}_j|\bm{X}_{i J(r,k)})$ if $j\in m(r)$ and $\hat{Z}_{ij}=X_{ij}$ otherwise,
	and $\bar{\varepsilon}_i = \varepsilon_i +  \{\bm{X}_{im(r)}-E(\bm{X}_{m(r)}|\bm{X}_{i J(r,k)})\} \bm{\beta}_{m(r)}^0$.
	\end{condition}

 \begin{condition}\label{C2}
 	For any $1\le r \le R$, with a sufficiently large $n$,
 	\begin{equation*}
 	P(\text{Columns of}\ \bm{G}^{(r)}(\bm{\beta}^0)\ \text{are linearly dependent} \text{ when } \bm{\beta}^0_{m(r)}\neq\bm{0})=0,
 	\end{equation*}
 	where $\bm{G}^{(r)}(\bm{\beta})=(\bm{g}_1^{(r)}(\bm{\beta}),\dots,\bm{g}_N^{(r)}(\bm{\beta}))^T$ is a submatrix of $\bm{G}(\bm{\beta})$ with columns representing estimating functions of the $r$-th group.
 \end{condition}




\begin{condition}\label{C4}
	For each $1\le j \le q$, there exists $1\le r\le R$ such that $j\in a(r)\cap \{\cup_{k\in \mathcal{G}(r)} a(k)\}$.
\end{condition}


	\begin{condition}\label{C5}
		There exists a covariance matrix $\bm{V}_1$ such that
		$\bm{G}^*_0(\bm{\beta}^0)^T\bm{1}/\sqrt{N}\overset{d}{\to}N(\bm{0}, \bm{V}_1)$, where  $\bm{G}_0^*(\bm{\beta})=\bm{G(\bm{\beta})\{\bm{U}(\bm{\beta}^0)\}^T}$ is the sample matrix for transformed estimating functions which are linearly independent at $\bm{\beta}=\bm{\beta}^0$.
\end{condition}

Note that, equations (\ref{imputeCondition1}) and (\ref{imputeCondition2}) in Condition \ref{C1} are satisfied if the consistency of a coefficient estimator for $\bm{\mathcal{X}}_{J(r,m)}$ holds under a linear model in predicting $X_l$, which can be obtained through the least squares or GLM \citep{fahrmeir1985consistency} estimator under MCAR or MAR mechanisms. 
Moreover, Condition \ref{C1} requires the existence of the fourth moments of covariates and the error term. Condition \ref{C2} holds when the block-wise imputations based on different missing pattern groups are distinct with probability $1$.
 Condition \ref{C4} is satisfied when the block-wise missing data contain complete cases, while for data with no complete cases, it requires that each relevant covariate is observed from at least one group and utilized in the MBI to predict missing values.
Since $\bm{G}(\bm{\beta})$ incorporates predicted conditional expectations, Condition \ref{C5} requires asymptotic normality of coefficient estimators for predicting missing covariates, and thus can be satisfied when the least squares estimator is used under the linear regression model and the missing mechanism is either MCAR or MAR. When there are no missing values, Condition \ref{C5} automatically holds.


To simplify expression of the following theorem, we define some notations. Let $\bm{\beta}_{A_1}$ and $\bm{\beta}_{A_2}$ be vectors of $\beta_j$ for $j\in A_1$ or $j\in A_2$, respectively. 
In addition, we let $\hat{\bm{V}}_2=\nabla_{A_1} (\bm{1}^T\bm{G}^*_0/N)(\bm{\beta}^0)$ be the first derivative of $\bm{1}^T\bm{G}^*_0/N$  with respect to $\bm{\beta}_{A_1}$, and $\bm{V}=  (\bm{V}_2 \bm{V}_1^{-1} \bm{V}_2^T)^{-1}$, where 
$\bm{V}_2$ is expectation of $\hat{\bm{V}}_2$.





	\begin{thm}\label{T1}
		Under Conditions \ref{C1}--\ref{C4}, if $\lambda_N \to 0$ and $\lambda_N\sqrt{N}/\zeta_N \to \infty$ as $N\to \infty$, then there exists a local minimizer $\hat{\bm{\beta}}$ of $f^*(\bm{\beta})$ such that the following properties hold:
		
		(i) Estimation consistency: $\|\hat{\bm{\beta}}-\bm{\beta}^0\|_2=O_p(N^{-1/2}\zeta_N)$.
		
		(ii) Sparsity recovery: $P(\hat{\bm{\beta}}_{A_2}=\bm{0})\to 1$ as $N\to \infty$.

		(iii) Asymptotic normality: If Conditions \ref{C5} hold, then $ \sqrt{N} (\hat{\bm{\beta}}_{A_1}-\bm{\beta}^0_{A_1}) \overset{d}{\to} N(\bm{0}, \bm{V})$.
		
	\end{thm}

Theorem \ref{T1} states that the proposed estimator is almost root-$n$ consistent and selects
the true model with probability approaching $1$. The convergence of the proposed estimator $\hat{\bm{\beta}}$ depends on the accuracy of predictions for the conditional expectations of missing covariates, since the proposed approach is based on imputation of the missing covariates. Thus, the $\zeta_N$ in (i) comes from the level of prediction accuracy assumed in Condition \ref{C1}. Note that if $\zeta_N$ is a constant, then the proposed estimator is root-$n$ consistent by Theorem \ref{T1}. In addition, the estimator of nonzero coefficients $\hat{\bm{\beta}}_{A_1}$ is asymptotically normal under Condition \ref{C5}. The empirical covariance matrix of $\hat{\bm{\beta}}_{A_1}$ is $\hat{\bm{V}}=(\hat{\bm{V}}_2  \hat{\bm{V}}_1^{-1} \hat{\bm{V}}_2^T)^{-1}$, where $\hat{\bm{V}}_1=\{\bm{G}^*_0(\bm{\beta}^0)\}^T\bm{G}^*_0(\bm{\beta}^0)/N$.




If only a single regression imputation based on complete observations is utilized, the sample estimating functions are $\bm{G}_{(1)}(\bm{\beta})=\bm{G}(\bm{\beta})\{\bm{U}_{(1)}\}^T$, where $\bm{U}_{(1)}$ selects estimating functions corresponding to the single imputation. Then, the empirical covariance matrix of the estimator induced by $\bm{G}_{(1)}(\bm{\beta})$ is $\hat{\bm{V}}^{(1)}$, where $\hat{\bm{V}}^{(1)}$, $\hat{\bm{V}}^{(1)}_1$, and $\hat{\bm{V}}^{(1)}_2$, are similarly defined as $\hat{\bm{V}}$, $\hat{\bm{V}}_1$, and $\hat{\bm{V}}_2$, respectively, except that $\bm{G}_0^*(\bm{\beta}^0)$ is replaced by $\bm{G}_{(1)}(\bm{\beta}^0)$.


In the following proposition, we show that utilizing the MBI improves the empirical efficiency of the parameter estimation with a smaller asymptotic variance
 compared with a single imputation approach. 

\begin{prop}\label{P1}
	Under the conditions in Theorem \ref{T1}, $\hat{\bm{V}}^{(1)}-\hat{\bm{V}}$ is positive semi-definite. 
\end{prop}

Proposition \ref{P1} indicates that the proposed estimator with MBI gains efficiency through incorporating additional information from incomplete case groups. 
In the following, we will establish consistency of the proposed estimator for diverging $p$ and $q$. 

	\vspace{-3mm}
\subsection{Consistency for diverging $p$ and $q$}
In this subsection, we consider cases when $p$ and $q$ increase as $N$ increases, that is, $p=p_N$ and $q=q_N$. We assume that the number of sources does not diverge as $N$ goes to infinity.
Let 
$\bm{H}(\bm{\beta}) =  \left( \{\bm{1}^T \partial_1 \bm{G}(\bm{\beta})/N\}^T,\dots, \{\bm{1}^T \partial_p \bm{G}(\bm{\beta})/N\}^T\right)^T$, $\bm{H}_{A_1}(\bm{\beta})$ and $\bm{H}_{A_2}(\bm{\beta})$ be sub-matrices of $\bm{H}(\bm{\beta})$ consisting of rows 
corresponding to covariates indexed by $A_1$ and $A_2$, respectively, where $\partial_j \bm{G}(\bm{\beta})$ denotes the first derivative of $\bm{G}(\bm{\beta})$ with respect to $\beta_j$ for $1\le j\le p$.
We also let  $\widehat{\bm{W}}(\bm{\beta})=\{\bm{U}(\bm{\beta}^0)\}^T\{\bm{W}^*_0(\bm{\beta})\}^{-1} \bm{U}(\bm{\beta}^0)$ be an estimator of the weighting matrix for all estimating functions, 
$\widetilde{\bm{W}}(\bm{\beta}) = \bm{H}_{A_1}(\bm{\beta})  \widehat{\bm{W}}(\bm{\beta})\{\bm{H}_{A_1}(\bm{\beta})\}^T$, and
$\mathcal{B}_0=\{\bm{\beta} : \|\bm{\beta}-\bm{\beta}^0\|_{\infty}\le N^{-\kappa_0}\zeta_N\}$ be a neighborhood of $\bm{\beta}_0$ for some constant $\kappa_0$, where $\bm{W}^*_0(\bm{\beta})= \bm{G}^*_0(\bm{\beta})^T \bm{G}^*_0(\bm{\beta})/N$ and $\zeta_N$ is a sequence such that $1/\zeta_N=O(1)$ and $\zeta_N=o(\log N)$. Denote the minimum signal by $\beta_{\min}=\underset{j\in A_1}{\min}|\beta^0_j|$, and the smallest eigenvalue of a matrix by $\lambda_{\min}(\cdot)$. For simplicity, we write w.p.a.$1$ as shorthand for ``with probability approaching one.''
 Since the dimensions of $\bm{G}$ and $\bm{\beta}$ diverge as $N$ grows, we require the following regularity conditions.

\begin{condition}\label{C6}
	For $\bm{\beta}\in\mathcal{B}_0$, and some 
	positive constants $\kappa_4$ and $\kappa_3 < \min\{\kappa'_1/2-\kappa_0/2, \kappa'_1/4-\kappa_2/4\}$, 
	$\underset{i,j\in A_1}{\max} \|\{\partial_i \bm{G}(\bm{\beta})\}^T \partial_j \bm{G}(\bm{\beta})/N\|_{\infty}=O_p(N^{\kappa_3})$, 
	$ \underset{1\le j\le p_N}{\max}
	\|\{\partial_j \bm{G}(\bm{\beta})\}^T \bm{G}(\bm{\beta})/N\|_{\infty} = O_p(N^{\kappa_3})$, 

		$\| \widehat{\bm{W}}(\bm{\beta}) \|_{\infty}= O_p(N^{\kappa_3})$,  
 $\|\bm{H}(\bm{\beta})\|_{\infty}= O_p(N^{\kappa_3})$, $\lambda_{\min}(\widetilde{\bm{W}}(\bm{\beta}))> \kappa_4$ w.p.a.$1$,
	where  $\kappa'_1=\kappa_1-1/6$, $\kappa_1\in(1/6, 1/2]$ and $\kappa_0, \kappa_2\in(0,\kappa'_1]$ are constants.
	
	
\end{condition}

Condition \ref{C6} controls the norms of matrices related to the estimating function matrix $\bm{G}$ for $\bm{\beta}$ in a neighborhood of true coefficients, which is similar to the conditions in \citep[Theorem B.1 and Theorem B.2]{fan2014endogeneity}. 
In particular, the condition assumes a lower bound for eigenvalues of $\widetilde{\bm{W}}$ to ensure a strict local minimizer of the objective function $f^*(\bm{\beta})$. 
Let $\bm{T}_N (\bm{\beta})=\nabla^2_{21}L_N(\bm{\beta}) \{\nabla^2_{11}L_N(\bm{\beta})\}^{-1}$, where $L_N(\bm{\beta})$ is the first term in (\ref{FinalObject}) with $\bm{U}=\bm{U}(\bm{\beta}^0)$. Here, $\nabla^2_{21}L_N(\bm{\beta})$ is a sub-matrix of the Hessian matrix of $L_N(\bm{\beta})$ with rows and columns indexed by $A_2$ and $A_1$, respectively, while $\nabla^2_{11}L_N(\bm{\beta})$ is defined similarly 
with rows and columns both indexed by $A_1$. 


\begin{condition}\label{penalty}
	For some constant $\tau_1>\kappa_0+\kappa_2$, $p'_{\lambda_N}(\beta_{\min}/2)=O(N^{-\tau_1})$, $\lambda_N^{-1}=O(N^{\eta_2})$, and $\lambda_N^{-1} \|\bm{T}_N (\bm{\beta})\|_{\infty} \le \min \{\eta_1/p'_{\lambda_N}(\beta_{\min}/2), O_p(N^{\eta_2}) \}$, where $\eta_1\in (0,1)$ and $\eta_2\in (0,\kappa'_1-2\kappa_3)$ are constants.
\end{condition}


Condition \ref{penalty} is standard for the SCAD penalty \citep{fan2011nonconcave}, where $p'_{\lambda_N}(\beta_{\min}/2)=O(N^{-\tau_1})$ can be satisfied as long as $\beta_{\min}$ is large enough, since $p'_{\lambda_N}(\cdot)$ is decreasing. The requirement for $\bm{T}_N(\bm{\beta})$ is similar to the irrepresentable condition of the SCAD penalty under high-dimensionality \citep{fan2011nonconcave}, but is derived for the loss function based on estimating equations instead of the least squares loss. 
	Specifically, when there are no missing values and $L_N (\bm{\beta})$ is the least-square loss function, then $\nabla^2_{21}L_N(\bm{\beta})$ is the sample covariance matrix between the irrelevant and relevant covariates, $\nabla^2_{11}L_N(\bm{\beta})$ is the sample covariance matrix among relevant covariates, and thus the condition on $\bm{T}_N(\bm{\beta})$ is exactly the same as the irrepresentable condition.
\begin{condition}\label{C7}
	There exists a constant $\tau_2> \kappa_1-1/6$ 
	such that 
	for $\bm{\beta}\in\mathcal{B}_0$, $1\le r\le R$, $k\in \mathcal{G}(r)$, 
			\begin{equation}\label{ExpectationEstimator}
		\sum_{j\in a(k)}\left|\frac{1}{N}\sum\limits_{i=1}^N I(\xi_i=r) \hat{Z}_{ij}\left\{E(\bm{X}_{im(r)}|\bm{X}_{i J(r,k)})-\hat{E}(\bm{X}_{m(r)}|\bm{X}_{i J(r,k)})\right\} \bm{\beta}_{m(r)} \right| = O_p\left(\frac{\zeta_N}{N^{\tau_2}}\right),
		\end{equation}
\vspace{-15mm}
	
	\begin{equation}\label{ExpectationEstimator2}
	\sum_{l\in m(r)}\left|\frac{1}{N}\sum\limits_{i=1}^N I(\xi_i=r) \bar{\varepsilon}_i
	\left\{E(X_{il}|\bm{X}_{i J(r,k)})-\hat{E}(\bm{X}_{l}|\bm{X}_{i J(r,k)})\right\}
	\right| = O_p\left(\frac{\zeta_N}{N^{\tau_2}}\right),
	\end{equation}
	where 
	$\hat{Z}_{ij}=\hat{E}(\bm{X}_j|\bm{X}_{i J(r,k)})$ if $j\in m(r)$ and $\hat{Z}_{ij}=X_{ij}$ otherwise,
 and $\bar{\varepsilon}_i = \varepsilon_i +  \{\bm{X}_{im(r)}-E(\bm{X}_{m(r)}|\bm{X}_{i J(r,k)})\} \bm{\beta}_{m(r)}$. We also assume that $X_j$ and $E(X_l|\bm{X}_{J(r, k)})$ are sub-Gaussian distributed for $1\le j \le p$, $l\in m(r)$, $1\le r \le R$, and $k \in \mathcal{G}(r)$. In addition, $\max\limits_{1\le r \le R} |m(r)|=o(N^{1/6})$ and $\max\limits_{1\le r \le R} \|\bm{\beta}_{m(r)}\|_{\infty}=O(\zeta_N)$.

\end{condition}

Condition \ref{C7} is analogous to Condition \ref{C1}. Similar to equations (\ref{imputeCondition1}) and (\ref{imputeCondition2}), equations (\ref{ExpectationEstimator}) and (\ref{ExpectationEstimator2}) can be obtained through the Lasso \citep{zhang2008sparsity} or SCAD \citep{fan2011nonconcave} under linear regression models and the MCAR or MAR mechanisms in predicting missing covariates, assuming that the magnitude of true coefficients and the numbers of missing covariates across groups do not diverge too fast as $N\to \infty$. We also assume that the covariates in Condition \ref{C7} are sub-Gaussian, since we allow the dimension of covariates $p_N$ to increase exponentially in the following theorem.

\begin{thm}[Consistency under high-dimensionality]\label{HighConsistency}
	Under Conditions \ref{C2} and \ref{C6}--\ref{C7}, if $\log p_N=O(N^{1-2\kappa_1})$, $q_N=O(N^{\kappa_2})$ and $\beta_{\min}> N^{-\kappa_0}\log N$, 
	 then there exists a strict local minimizer $\hat{\bm{\beta}}$ of $f^*(\bm{\beta})$ in (\ref{FinalObject}) such that the following properties hold:
	
	(i) Estimation consistency: 
	 $\|\hat{\bm{\beta}}-\bm{\beta}^0\|_{\infty}=O_p(N^{-\kappa_0}\zeta_N)$.
	 
	(ii) Sparsity recovery: $P(\hat{\bm{\beta}}_{A_2}=\bm{0})\to 1$ as $N\to\infty$.
\end{thm}

Theorem \ref{HighConsistency} states that when the number of covariates grows exponentially, the proposed method still processes estimation consistency and recovers sparsity accurately under regularity conditions. That is, the proposed estimator selects the true model with probability tending to $1$. We provide the proofs of Theorems \ref{T1}--\ref{HighConsistency} and Proposition \ref{P1} in the supplementary material.

\vspace{-3mm}
\section{Simulation study}\label{SectionSim}

In this section, we provide simulation studies to compare the proposed method with existing model selection approaches for handling block-wise missing data, including complete case analysis with the SCAD penalty (CC-SCAD), the single imputation with SCAD penalty (SI-SCAD), the iSFS, the DISCOM, and the DISCOM with Huber's M-estimate (DISCOM-Huber). The simulation results show that the proposed method achieves higher model selection accuracy than other competing methods through fully utilizing information from incomplete samples. We simulate data from a linear model (\ref{Lmodel}) using $50$ replications, where $\bm{\varepsilon}\sim N(\bm{0}, \bm{I}_N)$ and each row of $\bm{X}$ is independent and identically distributed from a normal distribution with mean $\bm{0}$ and an exchangeable covariance matrix determined by a variance parameter $\sigma^2=1$ and a covariance parameter $\rho$. 
We also carry out simulations with binary covariates or an unstructured correlation matrix. The simulation results for the unstructured correlation matrix are provided in the supplementary material.


The proposed method is implemented based on Algorithm 1. The imputation in SI-SCAD is estimated in a similar fashion but only based on the complete case group. The minimization problem in CC-SCAD and SI-SCAD is solved through the coordinate descent algorithm. We utilize the Matlab codes in 
 \resizebox{47ex}{1.4ex}{\href{https://github.com/coderxiang/MachLearnScripts}{\ttfamily https://github.com/coderxiang/MachLearnScripts}} to calculate the iSFS estimator. The implementation of DISCOM and DISCOM-Huber is provide by \citep{yu2019optimal}. In addition, we tune the parameter $\lambda$ for the CC-SCAD, SI-SCAD, and iSFS via BIC. Following \citep{yu2019optimal}, the $\lambda$ in DISCOM and DISCOM-Huber is tuned by a validation set which consists of $n_v$ random samples from the complete observations. For the methods with the SCAD penalty, we choose $a=3.7$ \citep{fan2001variable}.

We calculate the false negative rate (FNR) representing the proportion of unselected relevant covariates and the false positive rate (FPR) representing the proportion of selected irrelevant covariates as follows: 
\vspace{-3mm}
\[
\frac{\sum_{j=1}^{p} I(\hat{\beta}_j = 0, \beta_j^0 \neq 0)}{\sum_{j=1}^{p} I(\beta_j^0 \neq 0)}\  \ \text{ and }\ \  \frac{\sum_{j=1}^{p} I(\hat{\beta}_j \neq 0, \beta_j^0 = 0)}{\sum_{j=1}^{p} I(\beta_j^0 = 0)}.
\vspace{-3mm}
\]
 We say that a method has better model selection performance if the overall false negative plus false positive rate (FNR$+$FPR) is smaller.  We compare all the methods under the following five settings where the relevant predictors in Source $k$ share the same signal strength $\beta_{sk}$ for $k=1,\dots, S$. In the first two settings, we assume missing at random and missing completely at random, respectively, while we assume informative missing in the third and fourth settings. In addition, we consider data with no complete observations in the last setting. Under each setting, we also provide the computational time (in seconds) of each method with a given tuning parameter.
 

 \noindent\textbf{Setting $\bm{1}$:} Let $N=700$, $p=40$, $q=14$, $R=4$, $S=4$, $n_1=30$, $n_2=n_3=220$, $n_4=230$, $n_v=10$, $p_1=p_2=p_3=12$, $p_4=4$, $(\beta_{s1}, \beta_{s2}, \beta_{s3}, \beta_{s4}) = (5, 6, 7, 8)$, and $\rho=0.4 \text{ or } 0.7$. Each of the first three sources contains four relevant covariates, and the last source contains two relevant covariates. 
 Samples are sequentially randomly assigned into the complete case group with probabilities proportional to $\exp(-a_i)$ for $1\le i \le N$, where $a_i=10(X_{i37}+\dots+X_{i40})$ and $X_{i37},\dots,X_{i40}$ are the four covariates from Source $4$ for the $i$-th sample. Otherwise, they are uniformly assigned to the other three groups, where Sources $1$--$3$ have the same missing structure as the three sources in Groups $2$--$4$ in Figure \ref{missingpattern}, and Source $4$ covariates are all observed. This assignment ensures that samples with higher $a_i$ are less likely to be assigned to Group $1$ of the complete cases.


Since $a_i$ depends on Source $4$ covariates which are observed across all the missing patterns, samples in Setting $1$ are missing at random. The proposed method outperforms other competing methods for different correlations even with a high missing rate ($95.7\%$) as we are able to extract more information from incomplete samples. Table \ref{S1} shows that the overall FNR$+$FPR of the proposed method is the lowest among all methods. For example, when $\rho=0.7$, the FNR$+$FPR of the proposed method is $0.481$, which is only $66.5\%$, $56.7\%$, $62.5\%$, $63.5\%$, and $51.5\%$ of the FNR$+$FPR of CC-SCAD, SI-SCAD, DISCOM, DISCOM-Huber, and iSFS, respectively. Note that the FNR$+$FPR of iSFS is the same for different $\rho$ since the iSFS always selects Source $4$ covariates. This is possibly due to the larger weight of Source $4$ when applying the iSFS approach, as covariates of Source $4$ are observed in all samples. 

	Moreover, we provide the values of the two terms in MBI-BIC$_{\lambda}$ in equation (\ref{MBI-BIC}) for $20$ increasing values of $\lambda$ in Table 10 
	in the supplementary material.
Although the values of the RSS term are much higher than the values of the second term $df_{\lambda}\cdot \log (N)$, the changes of the RSS term across different $\lambda$'s are comparable to the changes of the second term, which indicates that the second term is able to determine the tuning of $\lambda$ and thus can prevent overfitting. We also illustrate this with Figure 3
in the supplementary material, where the red vertical line marks the smallest MBI-BIC$_{\lambda}$. Since the smallest MBI-BIC$_{\lambda}$ and the smallest first term do not correspond to the same value of $\lambda$, the second term is effective in preventing overfitting. Note that the first term $N\cdot\log\left\{RSS(\hat{\bm{\beta}}_{\lambda})/N\right\}$ in MBI-BIC$_{\lambda}$ is not strictly increasing as $\lambda$ increases. This is possibly due to that the first term in the proposed objective function (\ref{FinalObject}) is not strictly increasing of the RSS.

We investigate the performance of the proposed method under high-dimensional situations in the following Setting $2$.

   \noindent\textbf{Setting $\bm{2}$:} Let $N=500$, $p=1000$, $q=20$, $R=4$, $S=3$, $n_1=180$, $n_2=120$, $n_3=n_4=100$, $n_v=50$, $p_1=75$, $p_2=100$, $p_3=825$, $(\beta_{s1}, \beta_{s2}, \beta_{s3}) = (6, 5, 4)$, and $\rho=0.5\ \text{or } 0.8$. Sources $1$, $2$,  and $3$ contain $6$, $6$, and $8$ relevant covariates, respectively. 
All the samples are uniformly assigned to the four groups, which have the same missing structure as Groups $1$--$4$ in Figure \ref{missingpattern}.

The proposed method is more powerful in variable selection than other methods under the high-dimensional situations, as its FNR$+$FPR is the smallest among all the methods, as indicated in Table \ref{S2}. In particular, the proposed method performs especially effectively when correlations among covariates are as strong as $0.8$, with FNR$+$FPR $=0.048$, much smaller than any FNR$+$FPR of other methods. This is possibly because the strong correlations improve imputations in MBI, which compensate the negative effect of highly correlated covariates on variable selection under high-dimensional settings [\citealp{zhao2006model}, \citealp{fan2011nonconcave}].


  \noindent\textbf{Setting $\bm{3}$:} We consider the missing not at random. Let $N=250$, $p=60$, $q=15$, $R=4$, $S=3$, $n_1=n_2=45$, $n_3=n_4=80$, $n_v=10$, $p_1=p_2=p_3=20$, $(\beta_{s1}, \beta_{s2}, \beta_{s3}) = (2.5, 3, 3.5)$, and $\rho=0.4, 0.6,  \text{or } 0.8$. Each source contains five relevant covariates. Here, missing group assignment of the samples is the same as in Setting $1$, except that there is no Source $4$ and $a_i=3(X_{i1}+\dots+X_{i5}+y_i)$, where $X_{i1},\dots,X_{i5}$ are the $i$-th sample of the five relevant covariates from Source $1$.
  
  In Setting $3$, the probability of a missing sample depends on missing covariates and the response variable, which leads to informative missingness and biased imputation based on the complete group in SI-SCAD. In contrast, the proposed method, incorporating additional imputed values through aggregating different missing patterns, is able to reduce the selection bias caused by missingness. For example, when $\rho=0.6$, the FNR$+$FPR of the proposed method is $0.402$, less than those of other methods. 
  Note that the FNRs of DISCOM and DISCOM-Huber are small since these two methods tend to over-select variables, consequently producing large FPRs. On the other hand, the CC-SCAD tends to select fewer variables due to insufficient numbers of complete observations, which leads to small FPR and large FNR.
  
 In the following Setting $4$, we consider binary covariates. We first simulate data from a multivariate normal distribution with correlation $\rho$ similarly as in previous settings, and then transform each covariate $X_j$ in Source $1$ to $\sign(X_j)$.
  
   \noindent\textbf{Setting $\bm{4}$:} Let $N=700$, $p=60$, $q=15$, $R=4$, $S=3$, $n_1=45$, $n_2=n_3=265$, $n_4=125$, $n_v=20$, $p_1=p_2=p_3=20$, $(\beta_{s1}, \beta_{s2}, \beta_{s3}) = (7,8,10  )$, and $\rho=0.4 \text{ or } 0.7$. Sources $1$, $2$, and $3$ contain $2$, $6$, and $7$ relevant covariates, respectively. Missing group assignment of the samples is the same as that in Setting $3$, except that $a_i=10y_i$ for $1\le i \le N$.
  
  In addition to FNR and FPR, we also calculate the mean-squared-error (MSE) of the estimators. Table \ref{S6} shows that the proposed method has the smallest FNR$+$FPR and MSE among all the methods under Setting $4$, indicating that the proposed method performs better than other competing methods in both variable selection and coefficient estimation under the situations with binary covariates. Note that although the CC-SCAD does not perform well in estimation due to informative missing, it is still able to select variables more accurately than all other methods except the proposed method, especially for relatively small $\rho$. Moreover, the DISCOM and DISCOM-Huber perform the worst in this setting, possibly because the DISCOM methods are based on covariances.

   \noindent\textbf{Setting $\bm{5}$:} We follow similarly as in Setting $3$, except that there is no complete case group and  $R=3$. Let $N=300$, $n_1=n_2=n_3=100$, $(\beta_{s1}, \beta_{s2}, \beta_{s3}) = (0.8, 1, 1.5)$, $\rho=0.5, 0.6, 0.7, \text{or } 0.8$. All the samples are uniformly assigned to the three missing groups. 

   

\begin{table}\centering
	\begin{tabular}{*{5}{c}|*{4}{c}}
		\hline
		&\multicolumn{4}{c|}{$\rho=0.4$}	&\multicolumn{4}{c}{$\rho=0.7$}	\\
		\cline{2-9}
		\raisebox{1.5ex}[0pt]{Method}
		&FNR	&FPR	&\textbf{FNR+FPR}	& Time			&FNR	&FPR	&\textbf{FNR+FPR}  & Time	\\
		\hline
		\textbf{Proposed method}	&0.128 & 0.408 & 0.536	 &12.086			&0.116 & 0.365 & 0.481 &8.781	\\
		CC-SCAD				&0.493 & 0.131 & 0.624  	&0.016			&0.621 & 0.103 & 0.723	&0.026 \\
		SI-SCAD				  &0.357 & 0.427 & 0.784	 &0.004			&0.385 & 0.464 & 0.849	&0.073	\\
		DISCOM				 &0.000 & 0.859 & 0.859		&0.031		 &0.000 & 0.770 & 0.770  &0.034	\\
		DISCOM-Huber	&0.000 & 0.854 & 0.854		&0.051			&0.000 & 0.758 & 0.758	&0.055 \\
		iSFS        			&0.857	&0.077	&0.934		 &0.426		 &0.857	&0.077	&0.934	&0.614	\\
		\hline
	\end{tabular}
	\caption{FNR, FPR, and FNR$+$FPR under Setting $1$. ``Time'' represents the computational time (in seconds) of each method for one simulation with a given tuning parameter.}\label{S1}
\end{table}

\begin{table}[H]\centering
	\resizebox{\textwidth}{!}{
		\begin{tabular}{*{5}{c}|*{4}{c}}
			\hline
			&\multicolumn{4}{c|}{$\rho=0.5$}	&\multicolumn{4}{c}{$\rho=0.8$}	\\
			\cline{2-9}
			\raisebox{1.5ex}[0pt]{Method}
										&FNR	&FPR	&\textbf{FNR+FPR}	& Time			&FNR	&FPR	&\textbf{FNR+FPR}  & Time	\\
			\hline
			\textbf{Proposed method}	&0.033	&0.001	&0.034	 & $9.033\times10^3$				&0.042	&0.006	&0.048 &$9.724\times10^3$	\\
			CC-SCAD				&0.187	&0.017	&0.204		&0.859			&0.655	&0.013	&0.668	& 2.012 \\
			SI-SCAD				  &0.390	&0.008	&0.398		&0.488		&0.480	&0.038	&0.518 	& 1.152	\\
			DISCOM				 &0.006	 &0.367	&0.373			&74.800		 &0.023 &0.487	&0.510  &110.900	\\
			DISCOM-Huber	&0.033	&0.317	&0.350		&103.600			&0.117	&0.373	&0.491	&143.500 \\
			iSFS        			&0.537	&0.074	&0.611			&16.350		 &0.496 &0.096	&0.592	&19.674	\\
			\hline
	\end{tabular}}
	\caption{FNR, FPR, and FNR$+$FPR under Setting $2$. ``Time'' represents the computational time (in seconds) of each method for one simulation with a given tuning parameter.}\label{S2}
	\vspace{-3mm}
\end{table}

%
%
%
%
%
%

\begin{table}[H]\centering
		\resizebox{\textwidth}{!}{
	\begin{tabular}{*{5}{c}|*{4}{c}|*{4}{c}}
		\hline
		&\multicolumn{4}{c|}{$\rho=0.4$}	&\multicolumn{4}{c|}{$\rho=0.6$}		&\multicolumn{4}{c}{$\rho=0.8$}\\
		\cline{2-13}
		\raisebox{1.5ex}[0pt]{Method}
		&FNR	&FPR	&\textbf{FNR+FPR} & Time	&FNR	&FPR	&\textbf{FNR+FPR} & Time	&FNR	&FPR	&\textbf{FNR+FPR} & Time\\
		\hline
		\textbf{Proposed}	&0.096	&0.298	&0.394
		&9.831	&0.105	&0.297	&0.402
		&8.324	&0.123	&0.294	&0.417	&4.445\\
		
		CC-SCAD	&0.351	&0.090	&0.440
		&0.023	&0.461	&0.082	&0.543
		&0.053	&0.613	&0.070	&0.683	&0.182\\
		
		SI-SCAD	&0.492	&0.308	&0.800
		&0.206	&0.455	&0.288	&0.743
		&0.123	&0.492	&0.280	&0.772	&0.118\\
		
		DISCOM	&0.000	&0.541	&0.541
		&0.043	&0.001	&0.541	&0.542
		&0.041	&0.015	&0.465	&0.480	&0.052\\
		
		DISCOM-H	&0.009	&0.509	&0.518
		&0.092	&0.069	&0.472	&0.541
		&0.094	&0.196	&0.380	&0.576	&0.094\\
		
		iSFS        &0.425	&0.289	&0.714
		&0.320	&0.440	&0.304	&0.744
		&0.506	&0.479	&0.285	&0.764 &0.871	\\
		\hline
	\end{tabular}}
\caption{FNR, FPR, and FNR$+$FPR under Setting $3$. ``Proposed'' stands for the proposed method. ``DISCOM-H'' stands for the DISCOM-Huber method.}\label{S3}
\vspace{-3mm}
\end{table}

\begin{table}[H]\centering
	\resizebox{\textwidth}{!}{
	\begin{tabular}{*{6}{c}|*{5}{c}}
		\hline
		&\multicolumn{5}{c|}{$\rho=0.4$}	&\multicolumn{5}{c}{$\rho=0.7$}\\
		\cline{2-11}
		\raisebox{1.5ex}[0pt]{Method}
		&FNR	&FPR	&\textbf{FNR$+$FPR} &MSE & Time	&FNR	&FPR	&\textbf{FNR$+$FPR} &MSE&Time\\
		\hline
		\textbf{Proposed method}	&0.024	 &0.160     &0.184 		&1.570 &14.662	&0.052	&0.151	&0.203	&2.449 & 17.772	\\
		CC-SCAD			   &0.368	&0.096	   &0.464	   &16.409 
		&0.033	&0.588	&0.056	&0.644 &33.813 & 0.037	\\
		SI-SCAD				 &0.264	  &0.454	 &0.718		 &15.291	
		&0.196  &0.319	&0.451	&0.770	&18.715	 & 0.181\\
		DISCOM				&0.000	 &0.726		&0.726		&2.560 
		&0.052	&0.000	&0.703	&0.703	&3.909  & 0.052	\\
		DISCOM-Huber	&0.000	 &0.916		&0.916		&39.614	
		&0.115  &0.005	&0.872	&0.877	&47.117		& 0.087\\
		iSFS        			&0.215	&0.331		&0.545		&9.758	
		&0.499	&0.184	&0.420	&0.604	&11.932	&0.650	\\
		\hline
	\end{tabular}
	}
	\caption{FNR, FPR, FNR$+$FPR, and MSE under Setting $4$.}\label{S6}
\end{table}

\begin{table}[H]\centering
	\resizebox{\textwidth}{!}{
	\begin{tabular}{*{3}{c}|*{2}{c}|*{2}{c}|*{2}{c}|*{2}{c}}
		\hline
		&\multicolumn{2}{c|}{FNR}	&\multicolumn{2}{c|}{FPR}		&\multicolumn{2}{c|}{\textbf{FNR+FPR}} &\multicolumn{2}{c|}{MSE}&\multicolumn{2}{c}{Time}\\
		\cline{2-11}
		\raisebox{1.5ex}[0pt]{$\rho$}
		&\textbf{Proposed} 	&iSFS		&\textbf{Proposed}	&iSFS			&\textbf{Proposed}	&iSFS	&\textbf{Proposed}	&iSFS	&\textbf{Proposed}	&iSFS	\\
		\hline
		0.5	&0.267		&0.395  	&0.233		&0.380			&0.500		&0.774	 &0.258		&0.255	&8.849	&0.154\\
		0.6	&0.228  	&0.405		&0.154		&0.380			&0.382		&0.785	 &0.174		&0.262	&11.657	&0.180\\
		0.7	&0.221		&0.428		&0.135		&0.377			&0.356		&0.805	 &0.148		&0.296	&8.556	&0.200\\
		0.8	&0.201		&0.435		&0.114		&0.370			&0.316		&0.804	 &0.120		&0.326	&9.462	&0.302\\
		\hline
	\end{tabular}
	}
\caption{FNR, FPR, FNR$+$FPR, and MSE under Setting $5$. ``Proposed'' stands for the proposed method.}\label{S4}
\vspace{-3mm}
\end{table}

The proposed method is capable of handling data with no complete observations. However, complete observations are required for CC-SCAD, SI-SCAD, DISCOM, and DISCOM-Huber. Thus, we only compare the proposed method with iSFS in this setting. The proposed method performs better than iSFS on both estimation and variable selection especially when the correlations among covariates are strong.
 Table \ref{S4} shows that the FNR, FPR, and MSE of the proposed method are less than those of iSFS, respectively, in most situations. 
Moreover, the FNR$+$FPR of the proposed method decreases as $\rho$ increases, indicating that incorporating correlation information among covariates plays an important role in imputation especially when there are no complete cases. 

	
	 \begin{table}\centering
		\resizebox{\textwidth}{!}{
		\begin{tabular}{*{3}{c}|*{2}{c}|*{2}{c}|*{2}{c}}
			\hline
			&\multicolumn{2}{c|}{FNR}	&\multicolumn{2}{c|}{FPR}		&\multicolumn{2}{c|}{\textbf{FNR+FPR}}	&\multicolumn{2}{c}{Time}\\
			\cline{2-9}
			\raisebox{1.5ex}[0pt]{$\rho$}
			&Proposed 	&Proposed\_2		&Proposed 	&Proposed\_2			&Proposed	&Proposed\_2	&Proposed	&Proposed\_2	\\
			\hline
			0.4 & 0.030 & 0.031 & 0.272 & 0.270 & 0.301 & 0.301	&	17.115	&	13.836	\\
			0.5 & 0.025 & 0.029 & 0.380 & 0.438 & 0.406 & 0.467	& 12.471	&	16.913\\
			0.6 & 0.021 & 0.058 & 0.420 & 0.423 & 0.441 & 0.481	& 15.230	&	13.548\\
			0.7 & 0.054 & 0.081 & 0.485 & 0.587 & 0.539 & 0.668	&	10.448	&	15.167\\
			0.8 & 0.050 & 0.099 & 0.465 & 0.539 & 0.515 & 0.638	&	9.413	&	14.270\\
			\hline
		\end{tabular}
		}
		\caption{FNR, FPR, FNR$+$FPR, and MSE under Setting $6$. ``Proposed'' stands for the proposed method. ``Proposed\_2'' stands for the proposed method without using principle components for a non-singular $\bm{W}$\hspace{-0.5mm}.}\label{S_add}
	\end{table}

	
	In the following, we conduct additional simulations without using the principle components when $\bm{W}$ is non-singular. That is, we compare the estimators from solving equations (\ref{nonsingular}) and (\ref{FinalObject}), respectively.
	
 \noindent\textbf{Setting $\bm{6}$:} We proceed similarly as in Setting $1$, except that $n_1=n_2=n_3=n_4=200$ and $(\beta_{s1}, \beta_{s2}, \beta_{s3}, \beta_{s4}) = (5, 6, 7, 8)$.
 
We summarize the results of Setting $\bm{6}$ in Table \ref{S_add}, where ``Proposed'' stands for the proposed method and ``Proposed\_2'' stands for the proposed method without using principle components for non-singular $\bm{W}$\hspace{-0.5mm}.
Note that these two approaches perform similarly in terms of FNR, FPR and FNR+FPR. In addition, the proposed method with principle components performs slightly better than that without the principle components for settings with a large correlation $\rho$. This is possibly due to the fact that $\bm{W}$ might be close to singular for a large $\rho$ even though it is non-singular.


\vspace{-3mm}
\section{Real data application}\label{Real}

	In this section, we apply the proposed method to the Alzheimer's Disease Neuroimaging Initiative (ADNI) study \citep{mueller2005alzheimer} and compare it with existing approaches. A primary goal of this study is to identify biomarkers which can track the progression of Alzheimer's Disease (AD). Since the cognitive score from the Mini-Mental State Examination (MMSE) \citep{folstein1975mini} can measure cognitive impairment and is a diagnostic indicator of Alzheimer's disease \citep{tombaugh1992mini}, we treat the MMSE as the response variable, and intend to select biomarkers from three complementary data sources: MRI, PET, and gene expression. 
	Note that the sparsity assumption of the proposed method might not be suitable for raw imaging data or imaging data at small scales since images have to show some visible atrophy for AD. However, the sparsity assumption is still reasonable for region of interest (ROI) level data.
		Thus, we apply the proposed method to the ROI level data in ADNI instead of the raw imaging data. For raw imaging data, the principal components analysis (PCA) related methods [\citealp{doi:10.1080/01621459.2016.1261710}, \citealp{guo2015spatially}, \citealp{shen2015spatially}] could be more applicable.
	
	The MRI is segmented and analyzed in FreeSurfer by the Center for Imaging of Neurodegenerative Diseases at the University of California, San Francisco. Quantitative variables extracted from the MRI are volumes, average cortical thickness, standard deviation in cortical thickness, and surface areas of different ROIs.
	Quantitative variables from the PET images are computed by the Jagust Lab at the University of California, Berkeley. The PET features represent standard uptake value ratios (SUVR) of different ROIs, where the SUVR is an indicator of metabolic activity of a specific region.
	Gene expression variables are extracted form blood samples by Bristol-Myers Squibb laboratories, and represent expression levels at different gene probes.
	
	The response variable MMSE may not be measured at the same day as the imaging data, as the examinations could be time-consuming. We utilize the ``visit code'' provided by the ADNI study to link the MMSE and image data, which ensures that the MMSE and imaging data are measured within the same month. 
		We mainly focus on the MMSE and quantitative variables from the MRI, PET, and gene expression in the second phase of the ADNI study (ADNI-2) at month $48$, where block-wise missingness emerges due to low-quality images, high cost of measurements, or patients' dropouts. 
	
	The aim of our real data analysis is to select biomarkers associated with the MMSE, which may be useful for the prediction of the MMSE or Alzheimer's disease  in the future.
	There are $267$ MRI features, $113$ PET features, and $49,386$ gene expression variables. To reduce the bias in MRI caused by differences of brain sizes,
	we normalize the ROI volumes, surface areas and cortical thicknesses through dividing them by the whole brain volume, the total surface area, and the mean cortical thickness of each subject, respectively [\citealp{zhou2014significance}, \citealp{doi:10.1177/0962280217748675}].
	We screen out $300$ features from the gene expression predictors through sure independence screening (SIS) \citep{fan2008sure}, and select subjects containing observations from at least two sources. For the SIS procedure, since there are missing values in the data, we calculate the marginal correlation between the MMSE response and each predictor using all available pairs of observations from them, and then select predictors with relatively higher marginal correlations according to the conventional SIS. In total, there are $680$ features and $212$ subjects in four groups with $69$ complete observations, that is, $p=680$, $N=212$, and $R=4$, where the four groups have the same missing pattern structure as Groups $1$--$4$ in Figure \ref{missingpattern}. As the missing rate of this dataset is about $68\%$, it is important to fully utilize incomplete observations, such as in the proposed method.
	
	To compare the performance of the proposed method with existing methods, we randomly split the data into a test set and a training set $100$ times. Specifically, each test set consists of $43$ samples ($20$\% of all samples) randomly selected from the complete observations. The remaining $169$ samples ($80$\% of all samples) constitute the training set with $26$ complete observations, corresponding to a $85$\% missing rate of the training set.
	For the DISCOM and DISCOM-Huber method, we generate a validation set consisting of $n_v=10$ random samples from the complete observations in the training set.

\begin{table}\centering
		\resizebox{\textwidth}{!}{
	\begin{tabular}{*{6}{c}|*{5}{c}}
		\hline
		&\multicolumn{5}{c|}{$85\%$ missing rate}	&\multicolumn{5}{c}{$90\%$ missing rate}\\
		\cline{2-11}
		\raisebox{1.5ex}[0pt]{Method}
		&NS	&Mean	&SD  &\textbf{RI-RMSE}& Time&NS	&Mean	&SD  &\textbf{RI-RMSE}& Time\\
		\hline
		\textbf{Proposed}	
		&41		&4.320	&0.577	  &-- &	$1.517\times10^3$		&39 	&4.384	&0.403	&-- & $1.981\times10^3$\\
		CC-SCAD			    &13		&5.472	&0.918	&21.1\%	 & 0.081	&9		&5.692	&1.038	&23.0\% & 0.040 \\
		SI-SCAD				 &51	&5.548	&1.046  &22.1\%	 &	0.129	&50		 &5.990	&0.896	&26.8\% & 0.117\\
		DISCOM				&392	&29.964	&4.123	&85.6\% & 7.587			&366	&29.824	&4.026  	&85.3\% & 7.098\\
		DISCOM-H	&290	&29.978	&3.528	&85.6\%	 &	9.536	&258	&29.597	&3.393	&85.2\% & 9.012\\
		iSFS        			&43		&18.058	&1.987	&76.1\%  & 7.880	&44 	&19.056 	&2.113	&77.0\% & 6.220\\
		\hline
	\end{tabular}
	}
	\caption{The ``NS'' represents the mean number of selected variables. The ``Mean'' and ``SD'' represent mean and standard deviation of RMSE based on $100$ replications, respectively. The ``RI-RMSE'' for any method is the relative improvement of the proposed method over the competing method for the ADNI data. }\label{RD}
	\vspace{-3mm}
\end{table}

	We calculate the prediction root-mean-squared error (RMSE) $\sqrt{T^{-1}\sum_{i=1}^{T}(\hat{y}_i-y_i)^2}$ for each test set corresponding to each method, where $T$ is the number of observations in the test set, $y_i$ is the $i$-th true response value in the test set, and $\hat{y}_i$ is the corresponding fitted value using the model based on the training data. We also calculate the relative improvement (RI-RMSE) of the proposed method over other methods in terms of mean RMSE based on the $100$ replications. Specifically, the RI-RMSE of any given method is the ratio of the difference between the mean RMSE of the given method and the proposed method, to the mean RMSE of the given method. To investigate data with a higher missing rate, we also randomly partition all the samples into $25\%$ test and $75\%$ training sets with 
	$n_v=5$ for $100$ times, and calculate the corresponding RMSE and RI-RMSE in a similar fashion. With this partition, each training set contains $16$ complete observations corresponding to a $90$\% missing rate.

	In general, the proposed method achieves higher variable selection and prediction accuracy than all other methods for the ADNI data due to incorporating correlation information from incomplete observations. Specifically, Table \ref{RD} shows that the mean RMSE of the proposed method is smaller than that of any other method under two missing rates, even though the proposed method selects fewer variables than most of the other methods, which implies that the proposed method selects variables more accurately. More precisely, the proposed method reduces the RMSE of any other method by more than $20\%$, according to the RI-RMSE. Moreover, the relative improvement is still substantial for the missing rate $90\%$, indicating that the proposed method is more effective than other methods even when the missing rate is quite high. In addition, the proposed method produces smaller standard deviation of the RMSE and thus is more stable than most other methods.
	The CC-SCAD only selects $13$ or $9$ variables since there are only $26$ or $16$ complete observations for $80\%$ or $85\%$ training sets, respectively. The DISCOM and DISCOM-Huber select more variables than other methods, 
	which is consistent with the simulation findings in Section \ref{SectionSim}.
	
	Table 9
	in the supplementary material provides the first NS variables most frequently selected by each method based on the $100$ training sets with $85\%$ missing rate, where NS is the mean number of variables selected by the corresponding method. 
	The $41$ variables selected by the proposed method contain $29$, $2$, and $10$ biomarkers from the MRI, PET, and gene expression, respectively, most of which are also selected by other methods. In particular, the two PET biomarkers (``{\ttfamily RIGHT\_LATERAL\_VENTRICLE}'' and ``{\ttfamily RIGHT\_CHOROID\_PLEXUS}'') are also selected by the  DISCOM, and DISCOM-Huber, which represent SUVRs of the right lateral ventricle and the right choroid plexus, respectively. Note that the lateral ventricle and choroid plexus are indeed related to the AD [\citealp{apostolova2012hippocampal}, 
	\citealp{krzyzanowska2012pathological}].
	
	In addition, the ``{\ttfamily ST29SV},'' ``{\ttfamily ST40TA},'' ``{\ttfamily ST60TS},'' and ``{\ttfamily 11723246\_s\_at}'' are not only selected by the proposed method, but also selected by the CC-SCAD, DISCOM, and DISCOM-Huber. The ``{\ttfamily ST29SV},'' ``{\ttfamily ST40TA},'' and ``{\ttfamily ST60TS}'' are MRI features and represent the volume of the left hippocampus, the average cortical thickness of the left middle temporal gyrus, and the standard deviation of cortical thickness of the left temporal pole, respectively,  which are all associated with the presence of AD [\citealp{galton2001differing}, \citealp{apostolova2012hippocampal}, \citealp{convit2000atrophy}, \citealp{arnold1994neuropathologic}].
	The ``{\ttfamily 11723246\_s\_at}'' from the gene expression source represents the secreted frizzled related protein 1 (SFRP1) gene, which is elevated in brains of individuals with AD [\citealp{esteve2019elevated}, \citealp{tang2020enhancing}].
	Furthermore, the ``{\ttfamily ST43TA}'' and ``{\ttfamily ST119TS}'' from the MRI source are only selected by the proposed method, representing the average cortical thickness of the left paracentral lobule and the standard deviation of cortical thickness of the right temporal pole, respectively. Note that the left paracentral lobule and the right temporal pole are also associated with AD [\citealp{yang2019study}, \citealp{kumfor2016right}]. 
	
	In summary, the proposed method produces smaller RMSE for prediction in test sets than other competing methods with fewer selected variables, indicating that our method achieves better performance in variable selection. Moreover, the biomarkers selected by the proposed method are indeed important and relevant to the response variable, which are also confirmed by medical studies.

\vspace{-3mm}
\section{Discussion}
In this paper, we propose the multiple block-wise imputation approach to solve the block-wise missing problem arising from multi-source data. The proposed method improves variable selection accuracy through incorporating more information about missing covariates from incomplete case groups.

The existing methods for missing data do not fully utilize the structure of block-wise missing data to impute missing values and select relevant covariates. In contrast, the proposed MBI estimates missing variables within a group based on other group information, including complete and incomplete subject groups as well,
where the complete subject group contains more observed variables, while incomplete groups incorporate more samples. 
Moreover, when integrating all the block-wise imputations and missing patterns, the proposed method imposes more weight on estimating functions from groups with either fewer missing values or more accurate imputation. 


We show that the proposed method outperforms existing competitive methods in numerical studies, even for informative missing. Specifically, the proposed method is more powerful in handling informative missing data since the MBI reduces selection bias through aggregating more samples across different missing pattern groups than a single regression imputation based on complete cases. 
In addition, we establish the asymptotic normality, estimation and variable selection consistency for the proposed estimator. We also show that the proposed estimator is asymptotically more efficient than the estimator with a single imputation based on the complete case group. 

Although the MBI creates multiple predictions for each missing value to account for uncertainty of imputation, the proposed method is quite different from multiple imputation [\citealp{rubin2004multiple}] which draws multiple imputed values from a distribution, and utilizes each completed dataset separately. It is possible that the proposed method can be combined with MI through drawing more imputed values from the conditional distribution of missing variables, instead of relying on conditional expectation. In general, the idea of the MBI is flexible and can also be utilized with other predictive models besides the GLM, e.g., machine learning techniques such as the classification and regression tree-based approach \citep{loh2016classification}. Moreover, we can allow the inverse probability weighting in the MBI to adjust for unequal sampling in the future.





\vspace{-3mm}
	\section*{Acknowledgments}
	The authors thank the editor, the associate editor, and reviewers for providing thoughtful comments and suggestions. This work is supported by NSF grants DMS 1821198 and DMS 1613190.

{\setstretch{1.2} \footnotesize
	\bibliography{my_bib}
}

\end{document}